\documentclass[12pt]{iopart}

\usepackage{iopams}
\usepackage{graphicx}
\usepackage{subfig}
\usepackage{color}
\usepackage{ulem}
\usepackage{color}
\usepackage{ulem}
\usepackage{hyperref}

\begin{document}
	
	\title{Magnetic island formation and rotation braking induced by low-Z impurity penetration in an EAST plasma}
	
	\author{Shiyong Zeng}
	\address{Department of Engineering and Applied Physics, University of Science and Technology of China, Hefei, Anhui 230026, China}
	
	\author{Ping Zhu}
	\address{International Joint Research Laboratory of Magnetic Confinement Fusion and Plasma Physics, State Key Laboratory of Advanced Electromagnetic Engineering and Technology, School of Electrical and Electronic Engineering, Huazhong University of Science and Technology, Wuhan, Hubei 430074, China}
	\address{Department of Engineering Physics, University of Wisconsin-Madison, Madison, Wisconsin 53706, USA}
	\ead{zhup@hust.edu.cn}
	
	\author{Ruijie Zhou}
	\address{Institute of Plasma Physics, Chinese Academy of Sciences, Hefei, Anhui 230031, China}
	
	\author{Ming Xu}
	\address{Institute of Plasma Physics, Chinese Academy of Sciences, Hefei, Anhui 230031, China}
	

	\title[Simulation of impurity penetration and induced radiation-driven TM growth]{}
	
	\newpage
	\begin{abstract}
		Recent observations of the successive formations of the $4/1,3/1$, and $2/1$ magnetic islands as well as the subsequent braking of the $2/1$ mode during a low-Z impurity penetration process in EAST experiments are well reproduced in our $3D$ resistive MHD simulations.
		The enhanced parallel current perturbation induced by impurity radiation predominately contributes to the tearing mode growth, and the $2/1$ island rotation is mainly damped by the impurity accumulation as results of the influence from high $n$ modes.
	\end{abstract}

	\section{Introduction}
	The impurity radiation is long believed to play critical roles in setting the upper limits of energy confinement and stable operation regimes of tokamaks \cite{Suttrop_1997}. The exact mechanism how the impurity radiation influence and govern the tokamak stability has remained a subject of continued interests.
	
	Experiments have found some correlations between the impurity radiation and the tearing mode (TM) growth, ASDEX-Upgrade shows the low temperature and localized impurity radiation inside the growing magnetic island during the current contraction phase \cite{Suttrop_1997}, and the similar connection between enhanced radiation and enlarged island width is observed in NSTX \cite{Delgado_Aparicio_2011}. Besides, EAST experiments demonstrate the $m=2$ ($m$ is the poloidal mode number) island formation when the radiative cooling exceeds the ohmic heating \cite{Xu_2017}. Those experimental observation contribute to supporting the impurity radiation as a driven mechanism of TM but without more detail, JET experiments show the TM onset induced by unstable shrinking (broadening) current profile as a consequence of temperature edge cooling (central hollowing) from radiation loss \cite{Pucella2021}. In particular, Rijnhuizen tokamak uses the extended Rutherford model \cite{Rutherford1985} with a radiation term added to account for the observed mode exponential growth \cite{Salzedas2002}, although the data fitting is well it still remains some key parameters unknown due to the difficulties in diagnostic technologies.

	The idea of thermal instability induced mode growth was first proposed in Ref.\cite{Rebut1985}, and a thermo-resistive TM model is developed later to describe the effect of impurity radiation on the magnetic island nonlinear evolution \cite{Gates2012,White2015,Teng_2018}, which predicts the island growth once the local radiative cooling exceeds the Ohmic heating in the island interior. However, the model assumes the presence of a pre-existing small island or linearly unstable equilibrium to initiate the seeding required for the nonlinear island growth.

	Recent EAST experiments observe the successive $4/1,3/1,2/1$ and $3/2$ island formation during the carbon impurity penetration from lower divertor into the plasma core region \cite{XuMing2022}.
	The $2/1$ magnetic island, which propagates in electron diamagnetic drift direction, can be locked after the redistribution of carbon impurity concentration and the island width can reach approximately $5cm$ from the electron cyclotron emission (ECE) measurement.
	In addition, the ``hysteresis effect'' between the impurity concentration and the $2/1$ mode growth is found, namely the island width increases (decreases) and the rotation velocity decreases (increases) following the enhanced (reduced) impurity concentration in a hysteresis cycle.

	In this work, we are able to use 3D resistive MHD code NIMROD to simulate the process of impurity penetration from plasma boundary into central region and intent to reproduce the main features observed in the experiment. We demonstrate how the impurity radiation excites the TM growth and slows down the island rotation, even in absence of any external error field.

	The remainder of the paper is organized as follows. Section \ref{Section:2} gives a brief introduction to the simulation model and setup. Section \ref{Section:3} presents the simulation results about the $4/1,3/1$, and $2/1$ magnetic island formation after the arrival of impurity radiation peak on the corresponding rational surfaces, where the radiation enhanced parallel current perturbation predominantly contributes to the TM growth. Section \ref{Section:4} reports the reproduced island rotation damping by the concentration of impurity and demonstrates the effects of higher toroidal harmonics.
	Section \ref{Section:5} concludes with a summary and discussion.

	\section{NIMROD/KPRAD model and simulation setup}
	\label{Section:2}
	This work is based on a single-fluid 3D resistive MHD model implemented in the NIMROD code \cite{Sovinec2004}, with a simplified module for impurity radiation adapted from the KPRAD code \cite{KPRAD}, and the equations are as follows \cite{Izzo2013}:	
	\begin{eqnarray}
		\rho \frac{d \vec{V}}{dt} = - \nabla p + \vec{J} \times \vec{B} + \nabla \cdot (\rho \nu \nabla \vec{V})
		\label{eq:momentum}
		\\
		\frac{d n_i}{d t} + n_i \nabla \cdot \vec{V} = \nabla \cdot (D \nabla n_i) + S_{ion/3-body}
		\label{eq:contiune2}
		\\
		\frac{d n_{Z,Z=0-10}}{d t} + n_Z \nabla \cdot \vec{V} = \nabla \cdot (D \nabla n_Z) + S_{ion/rec}
		\label{eq:contiune3}
		\\
		n_e \frac{d T_e}{d t} = (\gamma - 1)[n_e T_e \nabla \cdot \vec{V} + \nabla \cdot \vec{q_e} - Q_{loss}]
		\label{eq:temperature}
		\\
		\vec{q}_e = -n_e[\kappa_{\parallel} \hat{b} \hat{b} + \kappa_{\perp} (\mathcal{I} - \hat{b} \hat{b})] \cdot \nabla T_e
		\label{eq:heat_flux}
		\\
		\frac{\partial \vec{B}}{\partial t} = \nabla \times \left( \vec{V} \times \vec{B} \right) - \nabla \times \left(  \eta \vec{j} \right) 
		\label{eq:ohm}
	\end{eqnarray}	
	Here, $n_i$, $n_e$, and $n_Z$ are the main ion, electron, and impurity ion number density respectively, $\rho$, $\vec{V}$, $\vec{J}$, and $p$ the plasma mass density, velocity, current density, and pressure respectively, $T_e$ and $\vec{q}_e$ the electron temperature and heat flux respectively, $D$, $\nu$, $\eta$, and $\kappa_{\parallel} (\kappa_{\perp})$ the plasma diffusivity, kinematic viscosity, resistivity, and parallel (perpendicular) thermal conductivity respectively, $\gamma = 5/3$ the adiabatic index, $S_{ion/rec}$ the density source from ionization and recombination, $S_{ion/3-body}$ also includes contribution from 3-body recombination, $Q_{loss}$ the energy loss, $\vec{E} (\vec{B})$ the electric (magnetic) field, $\hat{b}=\vec{B}/B$, and $\mathcal{I}$ the unit dyadic tensor.
	The source term $S_{ion/rec}$ in equation (\ref{eq:contiune3}) is from the impurity ionization and recombination, and the impurity radiation is calculated in the energy loss term $Q_{loss}$ from the KPRAD module based on a coronal non-equilibrium model including ionization, recombination, bremsstrahlung, and line radiations.
	All particle species share a single temperature $T=T_e$, which assumes instant thermal equilibration between the plasma and the impurity species, and the plasma pressure $p=\sum n_j T$, where $j$ refers to particle species, includes impurity contributions. More details can be found in Appendix \ref{appendix}.

	\begin{table}
		\caption{\label{input} Key parameters in the simulation}
		\footnotesize
		\begin{tabular}{@{}llll}
			\br
			Parameter & Symbol & Value & Unit \\
			\mr
			Minor radius & $a$ & $0.45$ & $m$ \\
			Major radius & $R_0$ & $1.85$ & $m$ \\
			Plasma current & $I_p$ & $0.38$ & MA \\
			Toroidal magnetic field & $B_{t0}$ & $2.267$ & T\\
			Core value of safety factor & $q_0$ & $1.484$ & dimensionless \\
			Edge value of safety factor & $q_{95}$ & $9.768$ & dimensionless \\
			Core electron number density & $n_{e,core}$ & $2.3 \times 10^{19}$ & $m^{-3}$ \\
			Core electron temperature & $T_{e,core}$ & $2.959$ & $keV$\\
			Edge electron temperature & $T_{e,edge}$ & $3.875$ & $eV$\\
			Equilibrium velocity & $V_0$ & $0$ & $m/s$\\
			The core resistivity & $\eta_0$ & $6.3316 \times 10^{-11}$  & $\Omega \cdot m$  \\
			Kinematic viscosity & $\nu$ & $27$  & $m^2/s$  \\
			The core Lundquist number & $S_0$ & $6.413 \times 10^9$  & dimensionless \\
			Constant perpendicular thermal conductivity & $\kappa_{\perp}$ & $1$ & $m^2/s$ \\
			Constant parallel thermal conductivity & $\kappa_{\parallel}$ & $10^{10}$ & $m^2/s$ \\
			Diffusivity & $D$ & $10$ & $m^2/s$\\
			\br
		\end{tabular}\\
	\end{table}

\begin{figure}[ht]
		\begin{center}
			\includegraphics[width=0.8\textwidth,height=0.7\textheight]{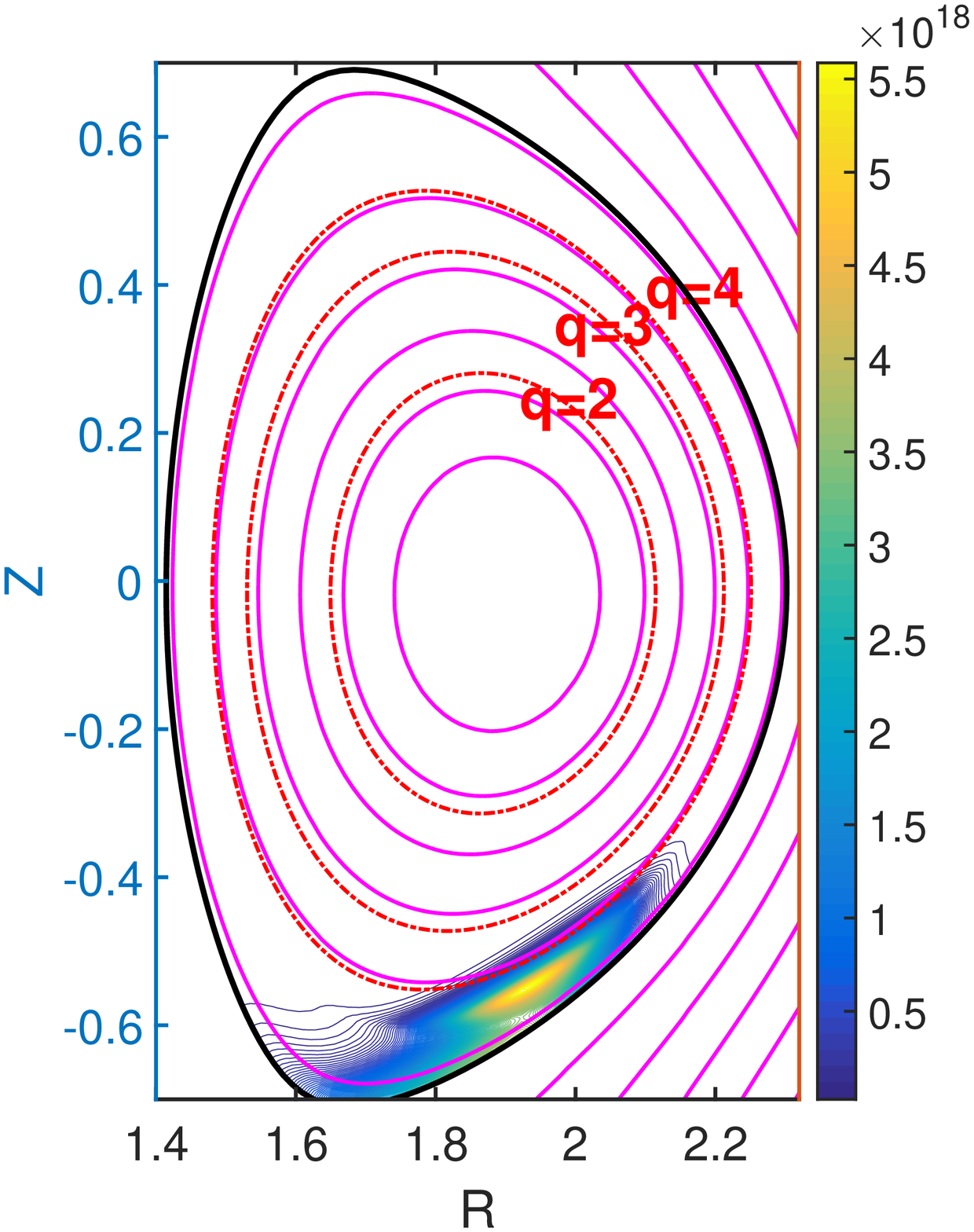}
		\end{center}
		\caption{Contours of the initial equilibrium $\psi$ (magenta solid lines) and impurity density distribution (flushed color, color bar in unit $m^{-3}$) in the poloidal plane, the equilibrium $q=4,3,2$ surfaces are denoted as red dashed lines and the boundary of simulation domain (plasma separatrix) is denoted as black solid line.}
		\label{psi_imp}
	\end{figure}

	A static MHD stable EAST L-mode equilibrium without any initial perturbations other than a localized impurity deposition is set up as the initial condition in simulations and some key parameters are listed in Table \ref{input}. 
	In the beginning of the simulation, we deposit an amount of neutral Neon impurity $8.73\times10^{18} m^{-3}$ at the bottom of the edge plasma region inside the plasma separatrix with a Gaussian distribution along both the poloidal and toroidal directions (Fig. \ref{psi_imp}).
	The impurity level is sufficiently low to avoid triggering the fast major disruption such as in a massive gas injection (MGI) process \cite{Zeng_2021}, while sufficiently high to excite the resistive tearing modes as in recent EAST experiments \cite{XuMing2022}.
	The impurity penetrates inward mainly through diffusion and convection as governed by equation (\ref{eq:contiune3}). A constant isotropic diffusion coefficient $ D = 10 m^2/s$ is adopted in the simulation, which is larger than the typical EAST experimental value, i.e. $0.1 \sim 1 m^2/s$, in order to accelerate the impurity inward penetration in simulations within the limit of affordable computational resource. The plasma and impurity are stationary at the beginning, i.e. $V_0 =0 m/s$, to exclude potential influence from the equilibrium plasma rotation and isolate the impurity effects on the magnetic island alone. Constant anisotropic thermal conductivities are used for simplicity, i.e. $\kappa_{\parallel} = 10^{10} m^2/s$ and $\kappa_{\perp} = 1 m^2/s$, and the temperature dependent Spitzer resistivity $\eta\propto T^{-3/2}_e$ is adopted. The plasma domain in simulation is limited by a perfectly conducting wall without a vacuum region.

	We use $70 \times 64$ bi-cubic Lagrange polynomial finite elements in the poloidal plane, and a semi-implicit time-advance is applied.
	Three simulations including different sets of toroidal mode numbers are studied in this work, i.e. $n=0$ only, $n=0-1$, and $n=0-5$. The $n=0-1$ case is able to reproduce the main observations from experiments, whereas the comparisons with other two cases show the influence from higher-$n$ modes.

	\section{The TMs driven by impurity radiation}
	\label{Section:3}
\begin{figure}[ht]
		\begin{center}
			\includegraphics[width=0.45\textwidth,height=0.3\textheight]{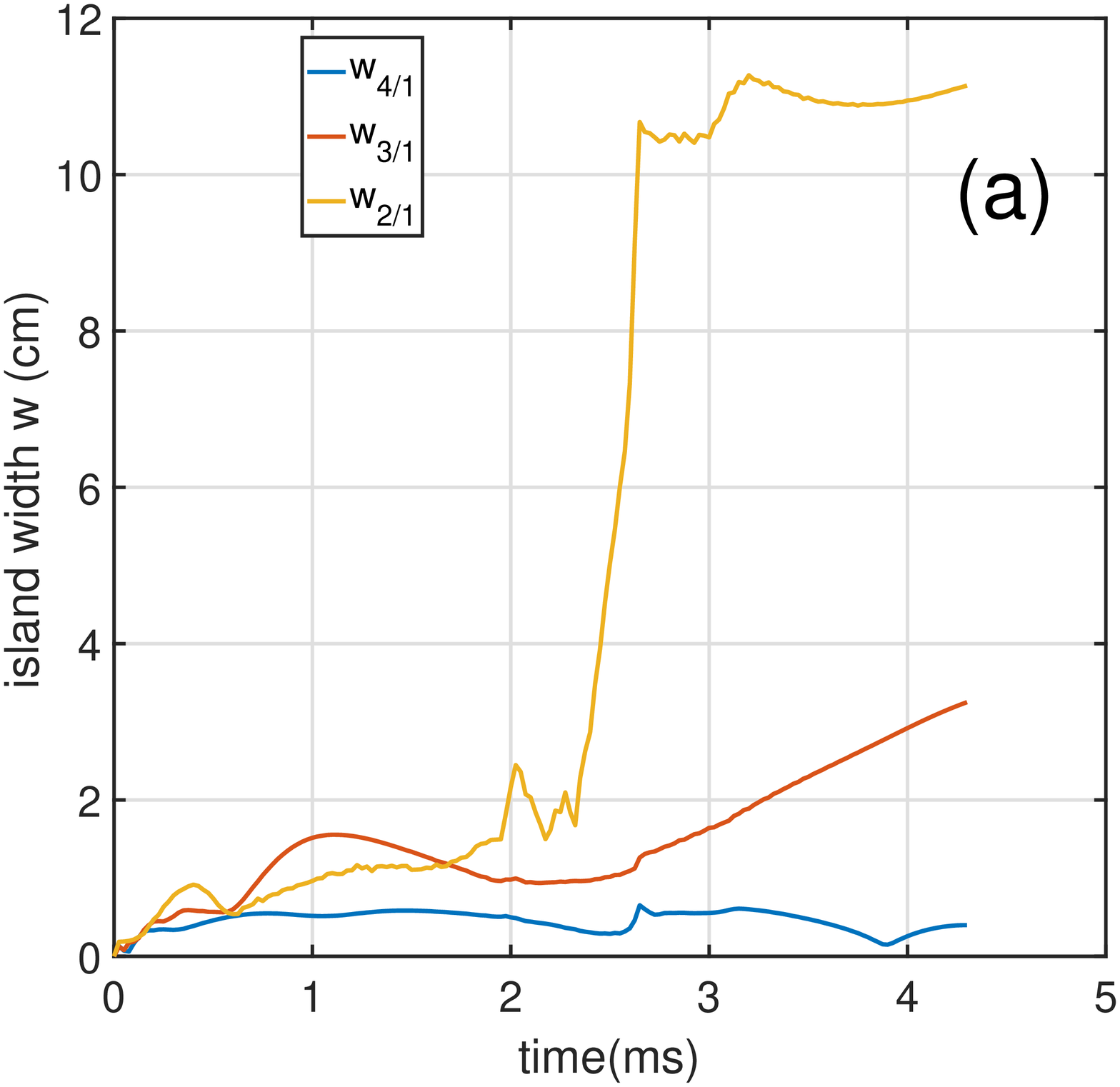}
			\includegraphics[width=0.45\textwidth,height=0.3\textheight]{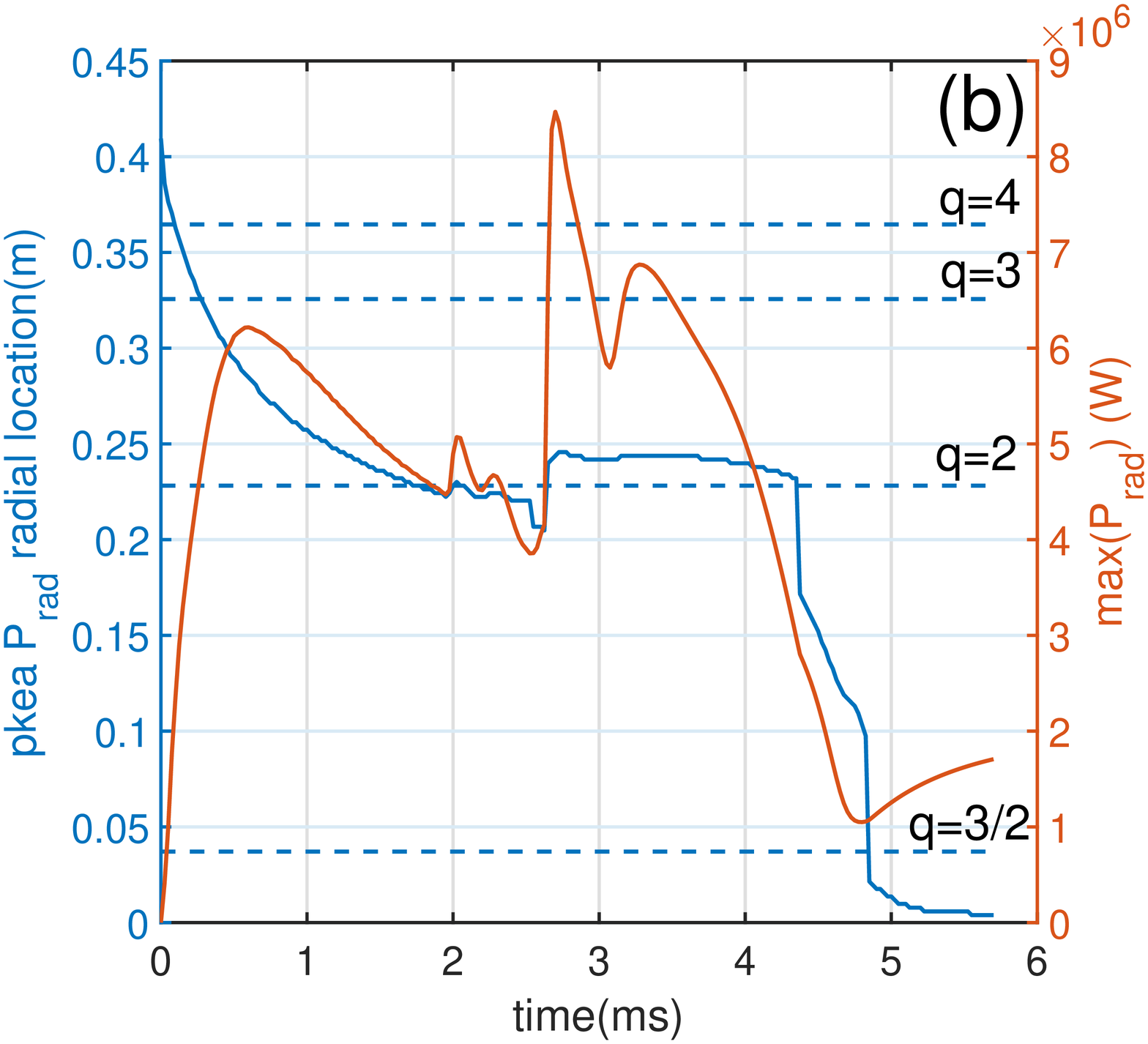}
			\includegraphics[width=0.85\textwidth,height=0.3\textheight]{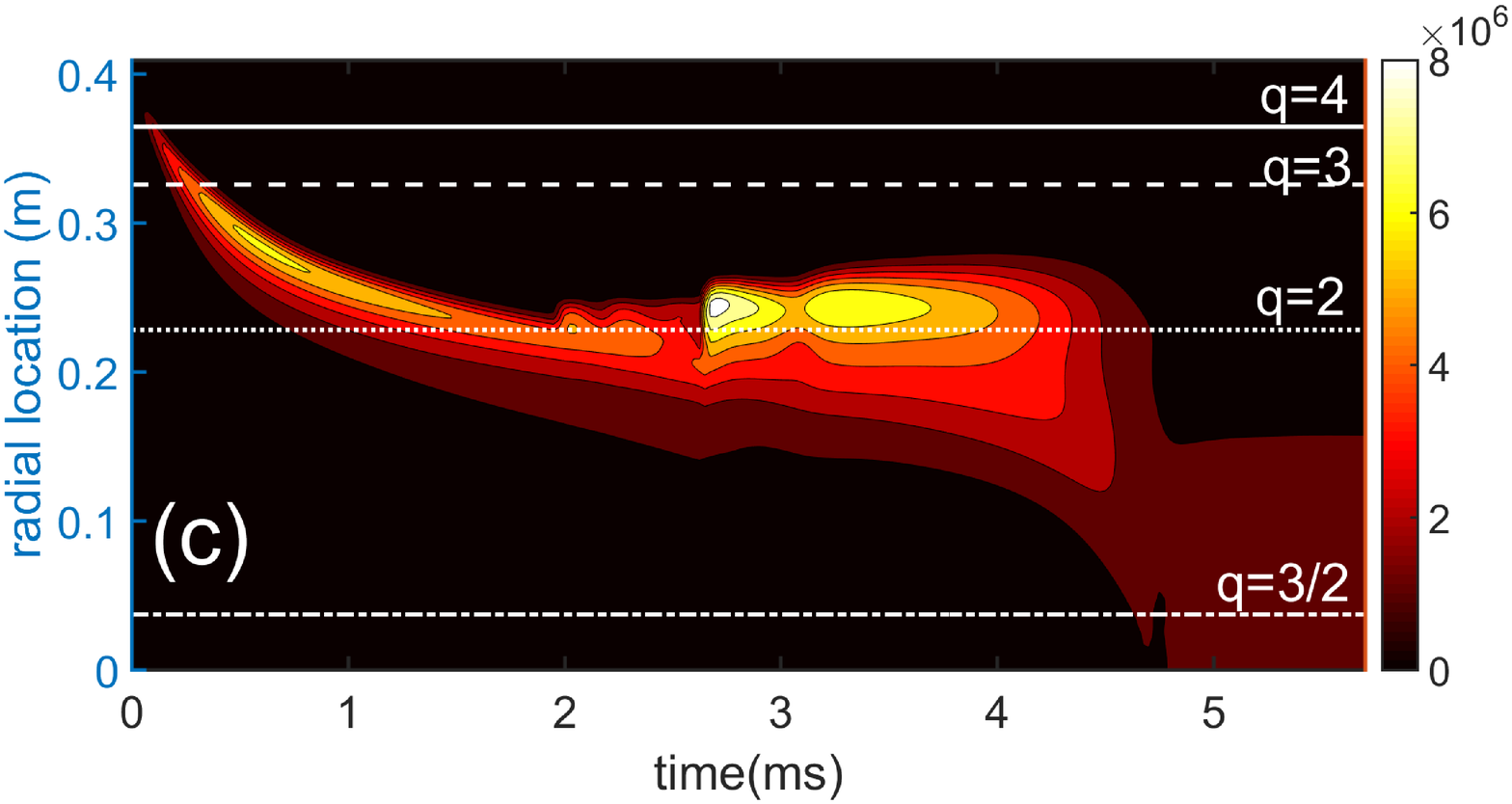}
		\end{center}
		\caption{(a) The island width of the $4/1$, $3/1$ and $2/1$ modes, (b) the radial location (blue solid line) and value (orange solid line) of impurity radiation power peak, and (c) the radial distribution of flux-surface-averaged impurity radiation power as a function of time, where the horizontal lines denote the radial location $r=R-R_0$ of equilibrium $q=4,3,2,3/2$ rational surfaces.}
		\label{imprad-tm}
	\end{figure}

	Our simulation results demonstrate the successive onsets and growth of the $4/1,3/1$, and $2/1$ islands during the impurity inward penetration (Fig. \ref{imprad-tm}a). The $4/1$ and $3/1$ island width increase and saturate first at $\sim 0.5cm$ and $\sim 2.0cm$ respectively, due to their proximity to the impurity source. The effect of impurity radiation on the island growth for the $2/1$ mode is more significant, as shown in its sudden fast growth at $t=2.5ms$ soon after the impurity radiation peak arrives on the $q=2$ rational surface at $t=2.0ms$ (Fig. \ref{imprad-tm}b), along with a strong burst of radiation power on the same surface (Figs. \ref{imprad-tm}b-c). The initial growth of the $2/1$ mode during $t=0-2.5ms$ is caused by mode coupling with the $3/1$ mode as shown in the following section.	
	Here the island width $w = 4(\frac{r_s q B_r}{m q' B_{p0}})^{1/2}$, where $B_{p0}$ is the equilibrium poloidal magnetic field and $B_r$ is the $m/n$ helical normal component of the perturbed magnetic field measured on the initial equilibrium rational surface $r_s$. The impurity radiation peak stops around the $q=2$ surface and does not penetrate inward further until after $t=4.5ms$. Similar process has been observed to trigger MHD instabilities in Tore Supra experiments \cite{Reux_2010}. Such a close correlation between the impurity penetration and the island growth indicates a crucial role of impurity radiation in triggering and driving the tearing mode.

	\subsection{Current perturbations induced by impurity radiation}
\begin{figure}[ht]
		\begin{center}
			\includegraphics[width=0.45\textwidth,height=0.3\textheight]{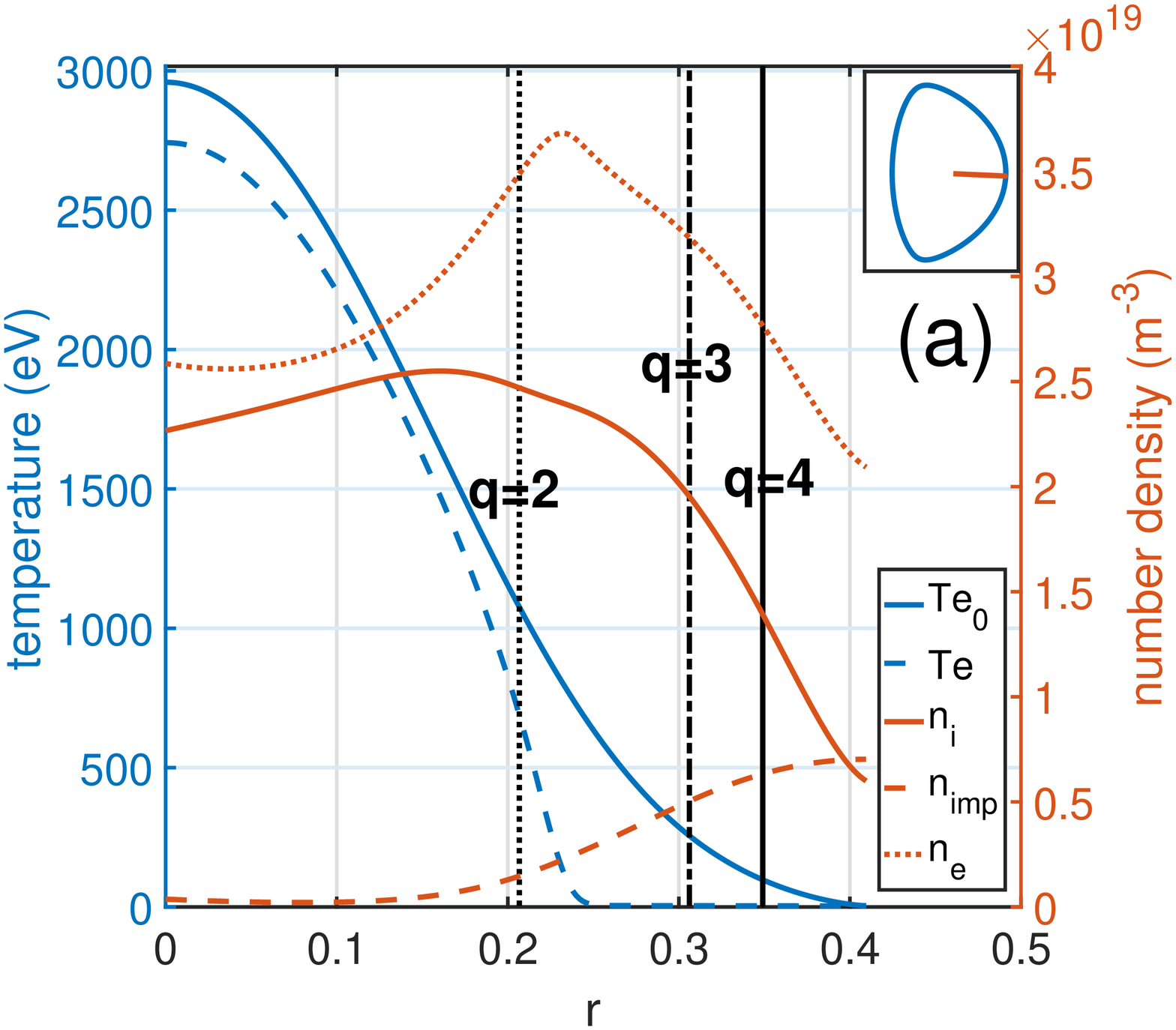}
			\includegraphics[width=0.45\textwidth,height=0.3\textheight]{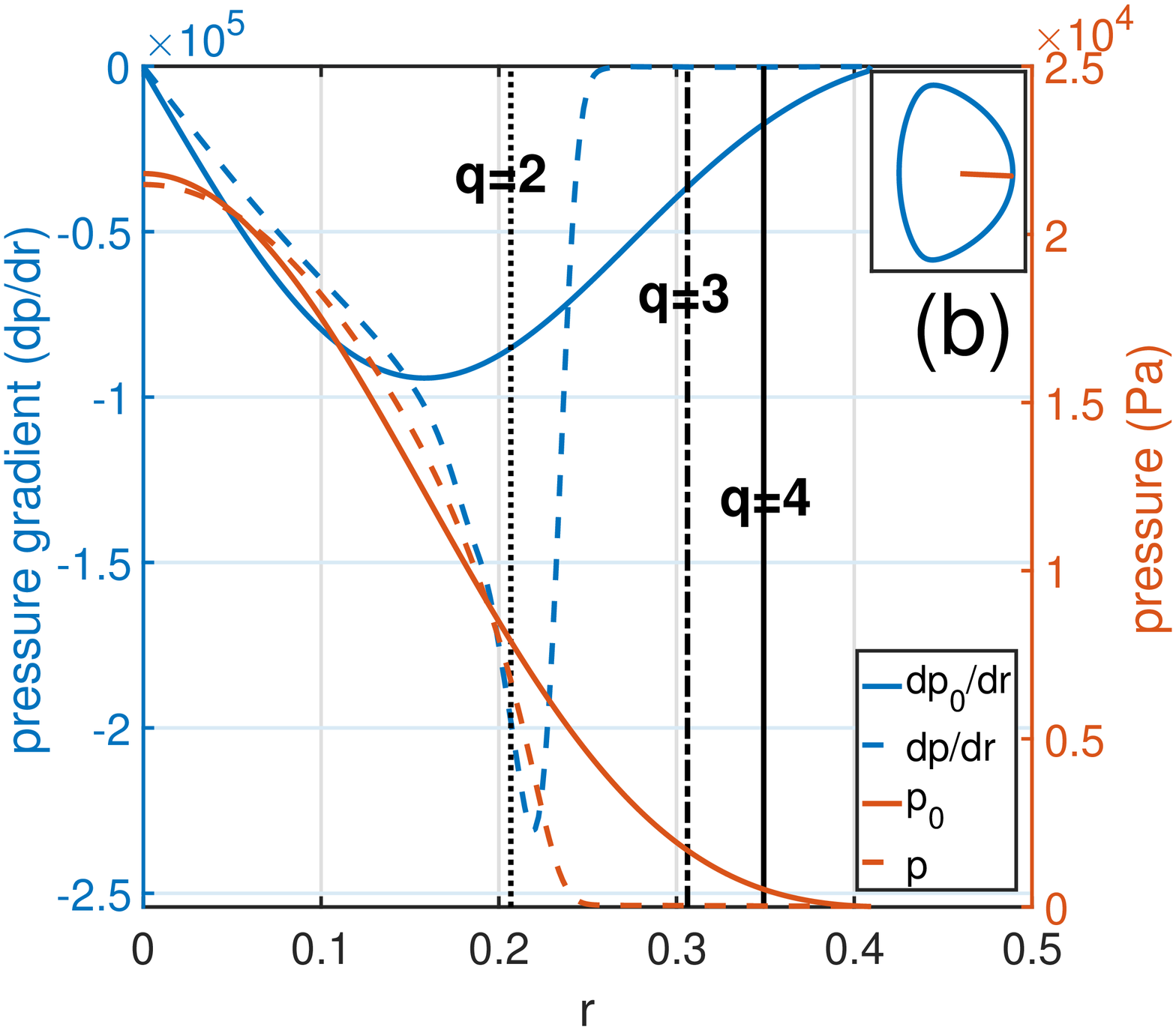}
			\includegraphics[width=0.45\textwidth,height=0.3\textheight]{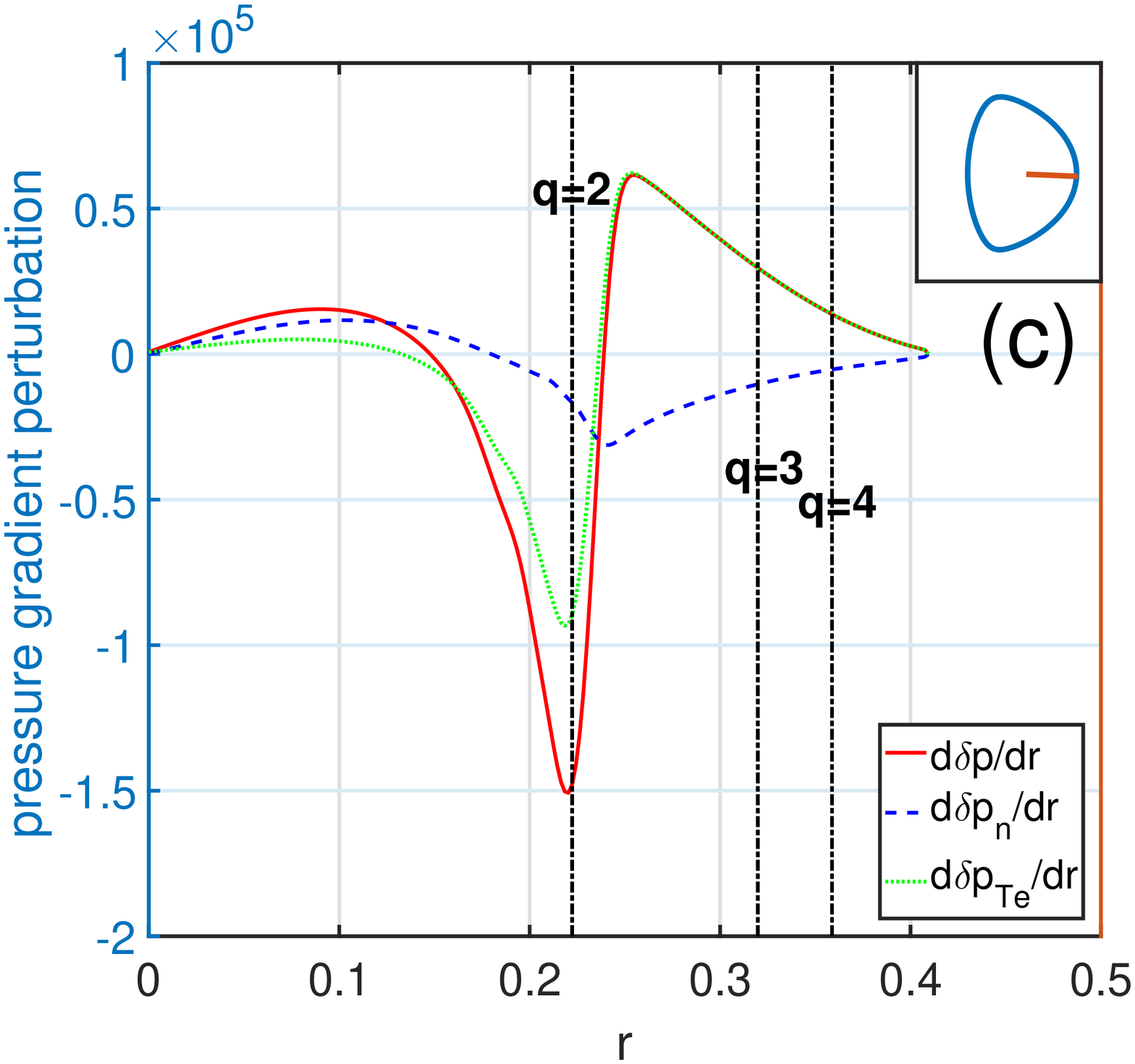}
		\end{center}
		\caption{Radial profiles along the outboard mid-plane at $t=1.5ms$ (with the radial line cut denoted as orange line in the inset sketch). Here $r=R-R_0$ and same in all other figures. (a) Left axis: initial equilibrium $T_{e0}$ (blue solid line) and dynamic $T_e$ (blue dashed line); right axis: plasma ion density $n_i$ (orange solid line), impurity ion density $n_{imp}$ (orange dashed line) and electron density $n_e$ (orange dotted line). (b) Right axis: initial equilibrium $p_0$ (orange solid line) and dynamic $p$ (orange dashed line); left axis: initial equilibrium pressure gradient $dp_0/dr$ (blue solid line) and dynamic pressure gradient $dp/dr$ (blue dashed line). (c) Pressure gradient perturbation $d\delta p/dr$, where $ p=nT_e$ (red solid line), pressure gradient perturbation with dynamic density and equilibrium temperature $d\delta p_n/dr$, where $ p_n=nT_{e0}$ (blue dashed line), and pressure gradient perturbation with dynamic temperature and equilibrium density $d\delta p_{T_e}/dr$, where $ p_{T_e}=n_0T_e$ (green dotted line). The equilibrium $q=4,3,2$ surface locations are denoted as black lines in all plots and same in all other figures.}
		\label{pressure_profile}
	\end{figure}

	Impurity injection introduces two direct modifications to the equilibrium profiles (Fig. \ref{pressure_profile}). On the one hand, the impurity radiation cooling leads to the temperature profile contraction; on the other hand, the impurity ionization increases the electron density at the impurity cold front, which is identified from the local peak of the electron density profile (Fig. \ref{pressure_profile}a). As a consequence, the pressure profile steepens and its gradient is enhanced at the impurity cold front around the $q=2$ surface (Fig. \ref{pressure_profile}b). This local variation in pressure gradient is predominantly from the temperature profile change in the simulation case considered here.

\begin{figure}[ht]
		\begin{center}
			\includegraphics[width=0.7\textwidth,height=0.42\textheight]{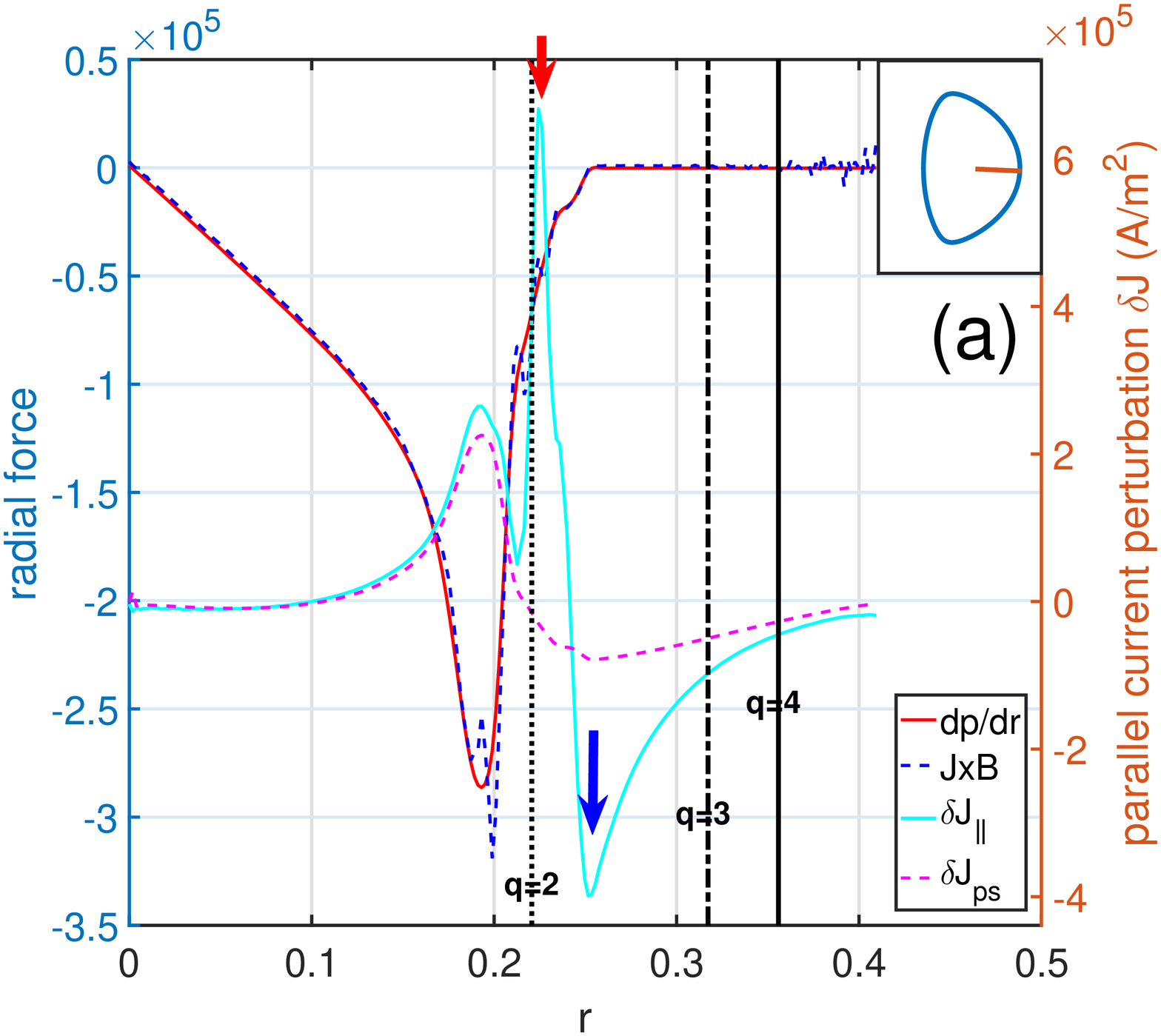}
			\includegraphics[width=0.7\textwidth,height=0.42\textheight]{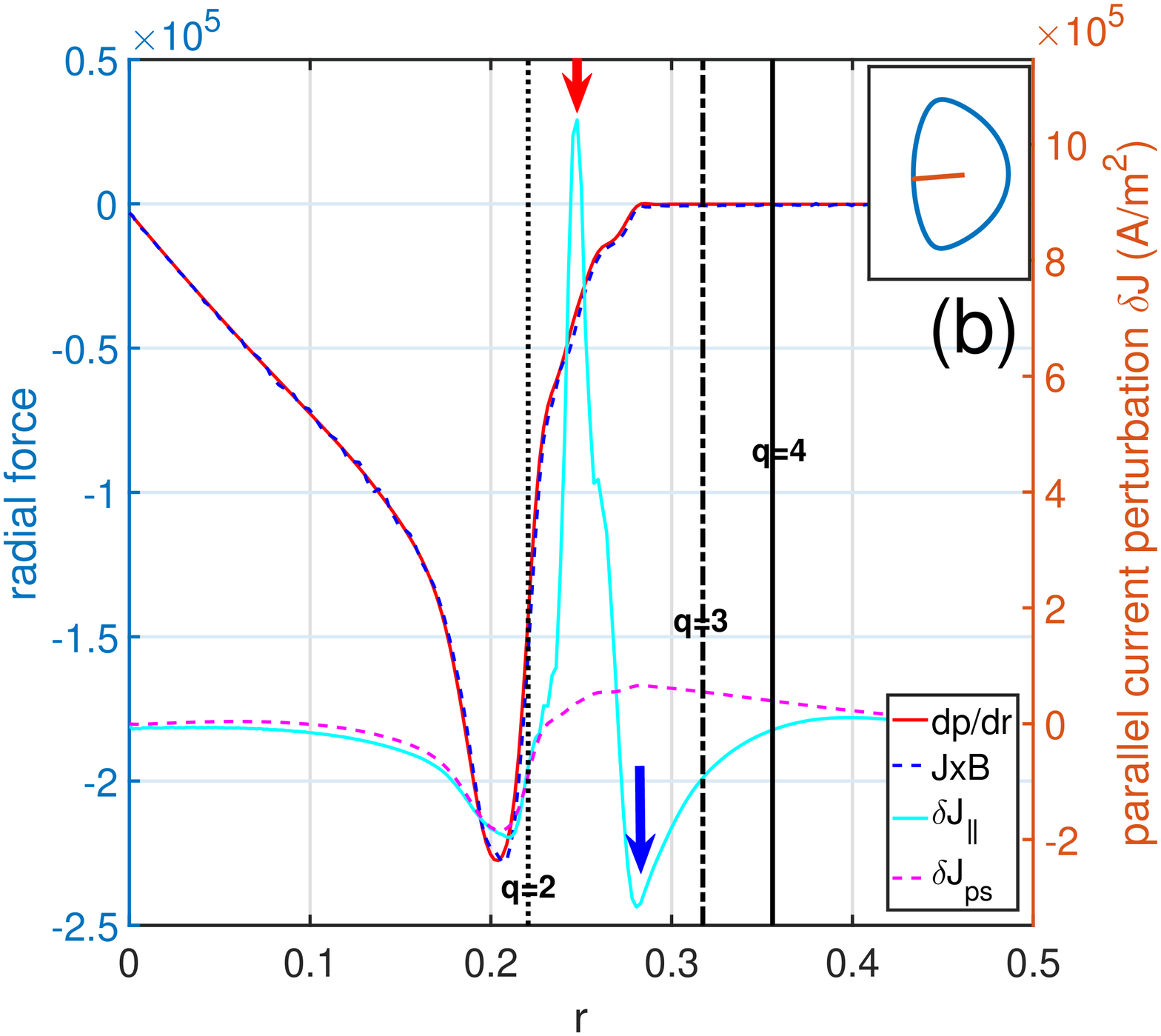}
		\end{center}
		\caption{Radial profiles along the (a) outboard and the (b) inboard mid-plane (with the radial line cut denoted as orange line in the inset sketch) for (left axis) pressure gradient $dp/dr$ (red solid line) and radial Lorentz force $\left( J \times B\right)_r = J_{\theta} \times B_{\phi} - J_{\phi} \times B_{\theta}$ (blue dashed line), (right axis) perturbed parallel current density $\delta J_{\parallel}$ (cyan solid line), and perturbed Pfirsch-Schl\"{u}ter current model $\delta J_{ps}$ (magenta dashed line).}
		\label{p-s current}
	\end{figure}

	Despite the dynamic nature of the impurity injection process, the profile evolution is quasi-static and the force balance $dp/dr \approx J \times B$ is well maintained (Fig. \ref{p-s current}).
	The corresponding parallel current perturbation $\delta J_{\parallel}$ mainly comes from the perturbed Pfirsch-Schl\"{u}ter (PS) current $\delta J_{ps} = -2 \frac{1}{B_p} \frac{r}{R} \frac{d\delta p}{dr} \cos\theta$ in the outer region on both sides of the $q=2$ surface, which is induced by the pressure gradient perturbation $d\delta p/dr$.
	However, the parallel current perturbation peaks around the $q=2$ surface can not be accounted by the perturbed Pfirsch-Schl\"{u}ter current $\delta J_{ps}$ alone.

\begin{figure}[ht]
		\begin{center}
			\includegraphics[width=0.7\textwidth,height=0.42\textheight]{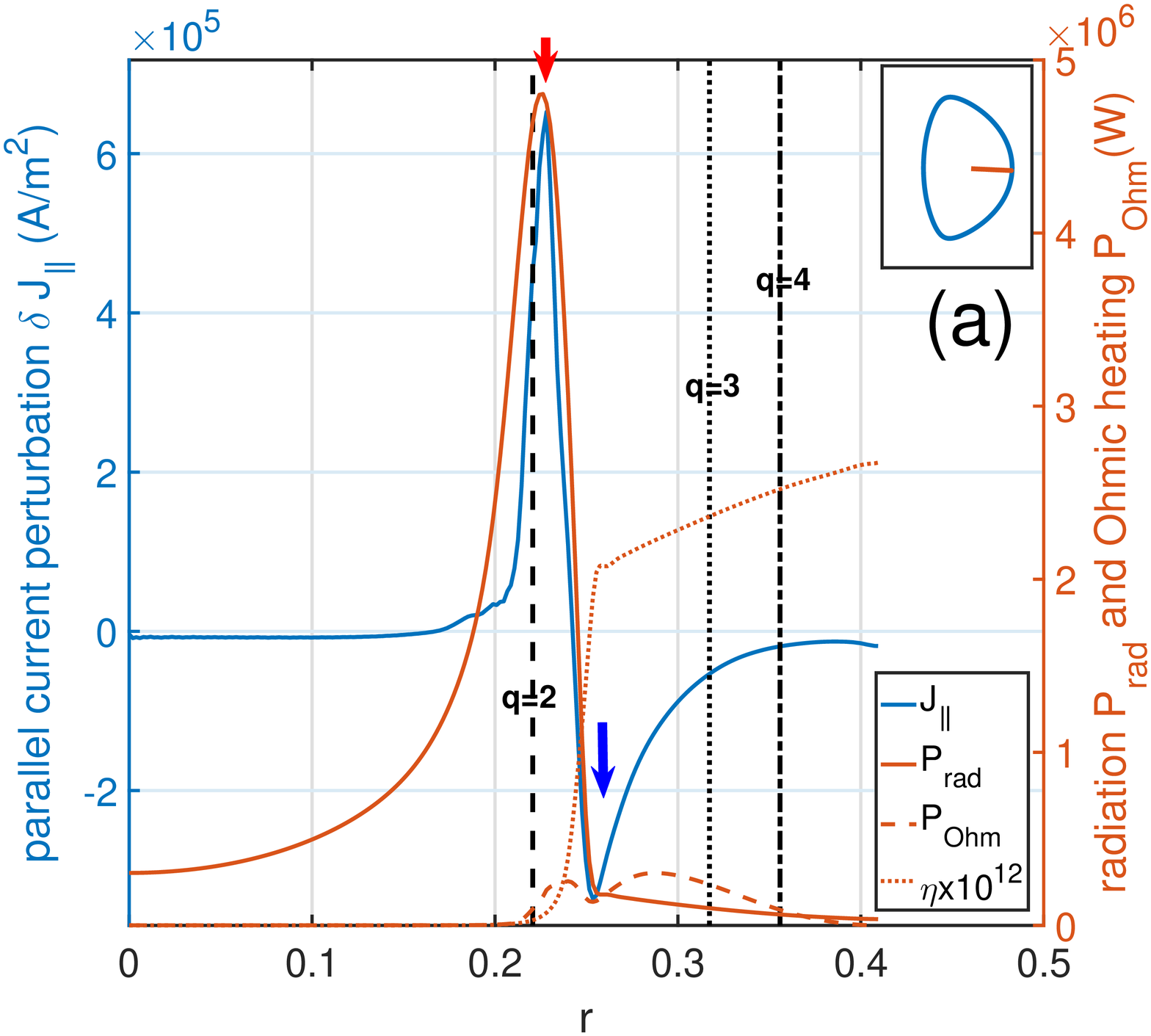}
			\includegraphics[width=0.7\textwidth,height=0.42\textheight]{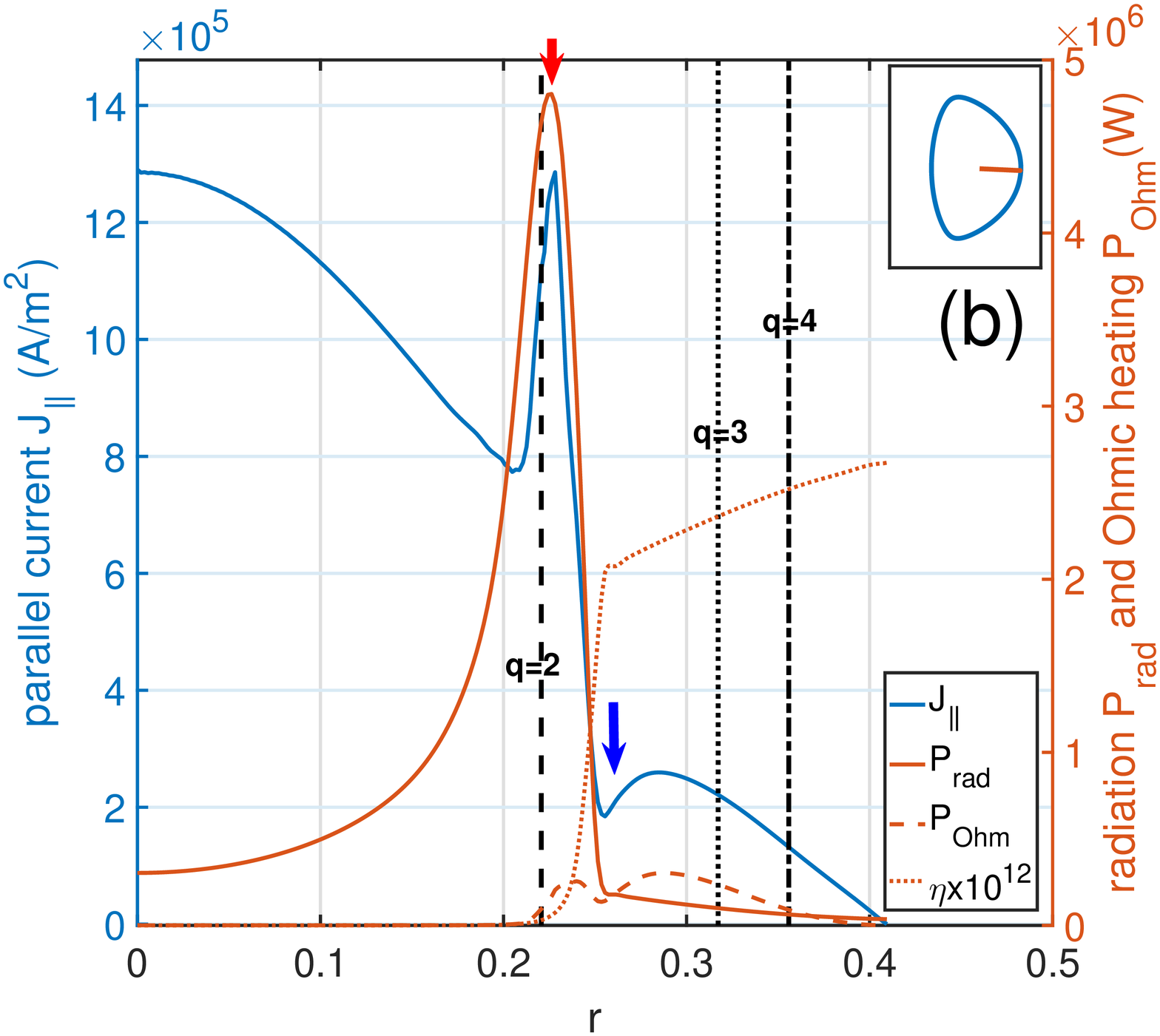}
		\end{center}
		\caption{Flux-surface-averaged profiles of impurity radiation power $P_{rad}$ (orange solid line), Ohmic heating power $P_{Ohm}$ (orange dashed line), plasma resistivity $\eta$ (orange dotted line), and (a) parallel current perturbation $\delta J_{\parallel}$ (blue solid line), (b) parallel current $J_{\parallel}$ (blue solid line).}
		\label{radiation-current}
	\end{figure}

	The local positive current perturbation peak denoted by the red arrow correlates well with the local impurity radiation peak (Fig. \ref{radiation-current}).
	Similar skin current structure due to edge radiative cooling is also reported in previous studies \cite{Robert1990,Ferraro_2018,Nardon_2021}.
	The local current perturbation dip denoted by the blue arrow collocates with the surge of plasma resistivity in the cold region and is dominated by the current diffusion (Fig. \ref{radiation-current}). It is important to note that the location of impurity radiation peak is usually different from that of the enhanced plasma resistivity, because the impurity radiation is dominated by the line radiation, which is the strongest around $T_e \sim 10^2 eV$, whereas the resistivity is proportional to $T_e^{-3/2}$ and more enhanced in the colder region.

	\subsection{Correlation between the current perturbations and TMs}
\begin{figure}[ht]
		\begin{center}
			\includegraphics[width=0.7\textwidth,height=0.42\textheight]{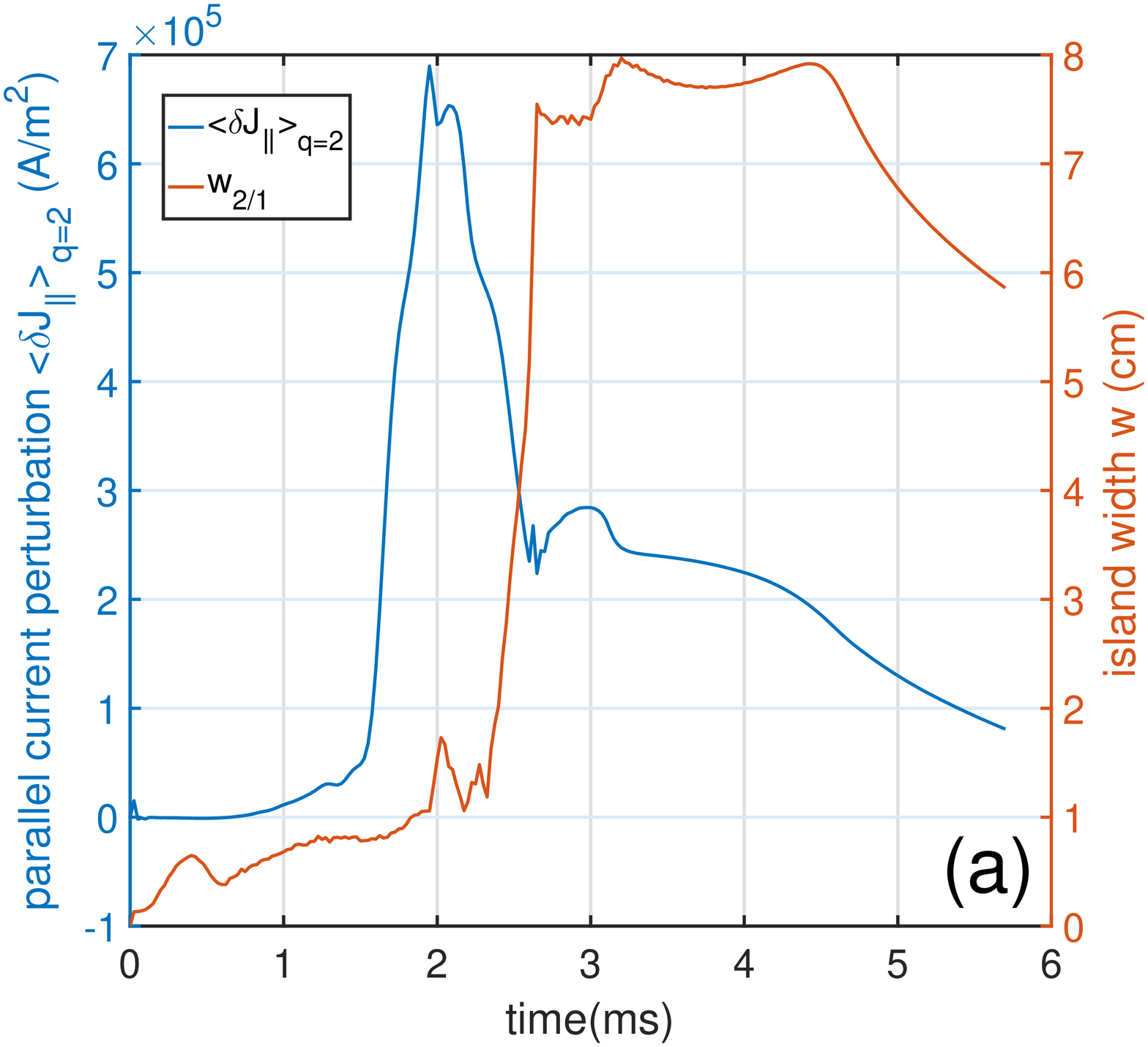}
			\includegraphics[width=0.7\textwidth,height=0.42\textheight]{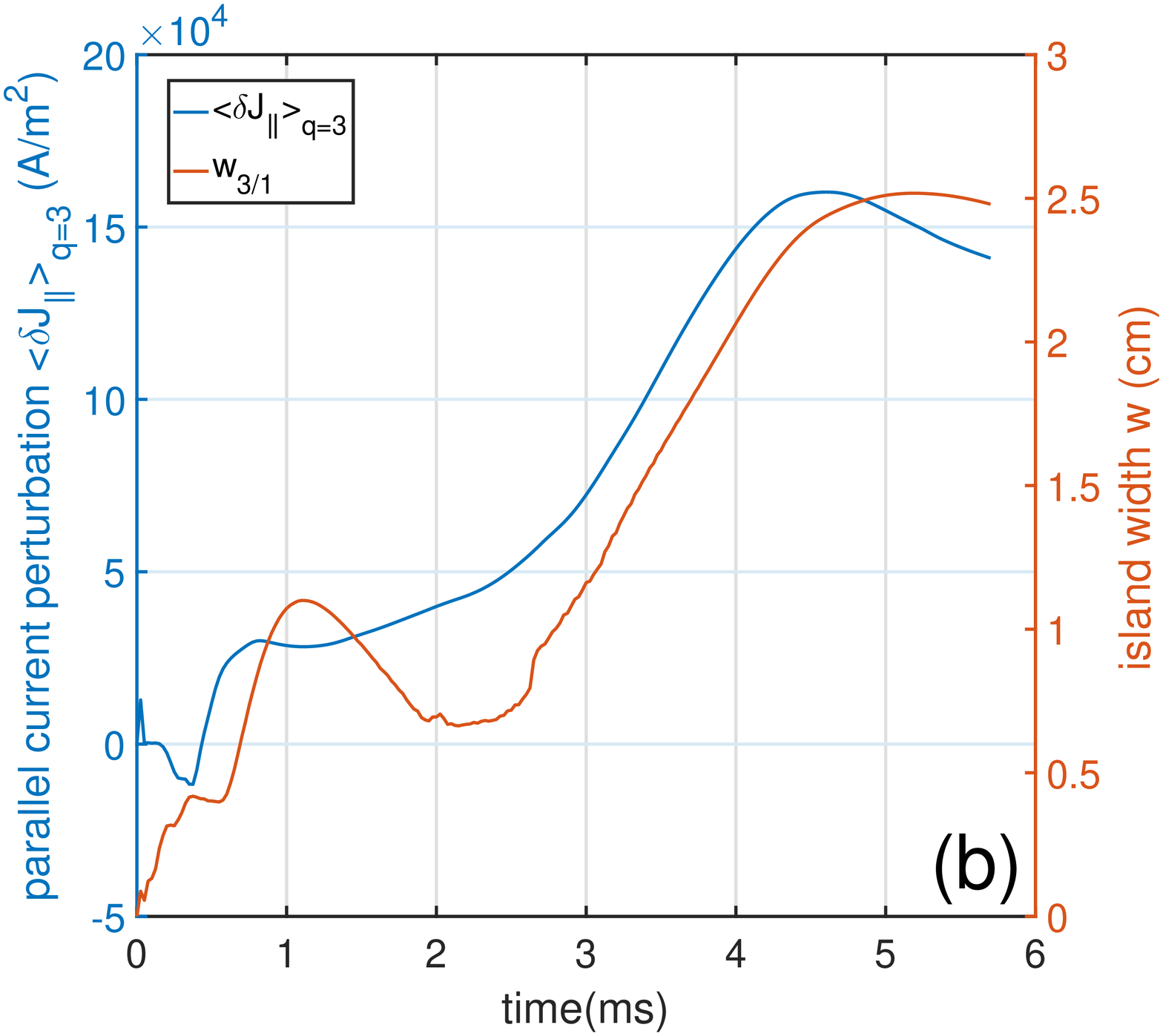}
		\end{center}
		\caption{Island width $w$ (orange solid line) and rational-surface-averaged parallel current perturbation $\left\langle \delta J_{\parallel}\right\rangle_{q=m/n}$ (blue solid lone) for (a) the $2/1$ mode and (b) the $3/1$ mode as a function of time.}
		\label{deltaj-width}
	\end{figure}

\begin{figure}[ht]
		\begin{center}
			\includegraphics[width=0.7\textwidth,height=0.42\textheight]{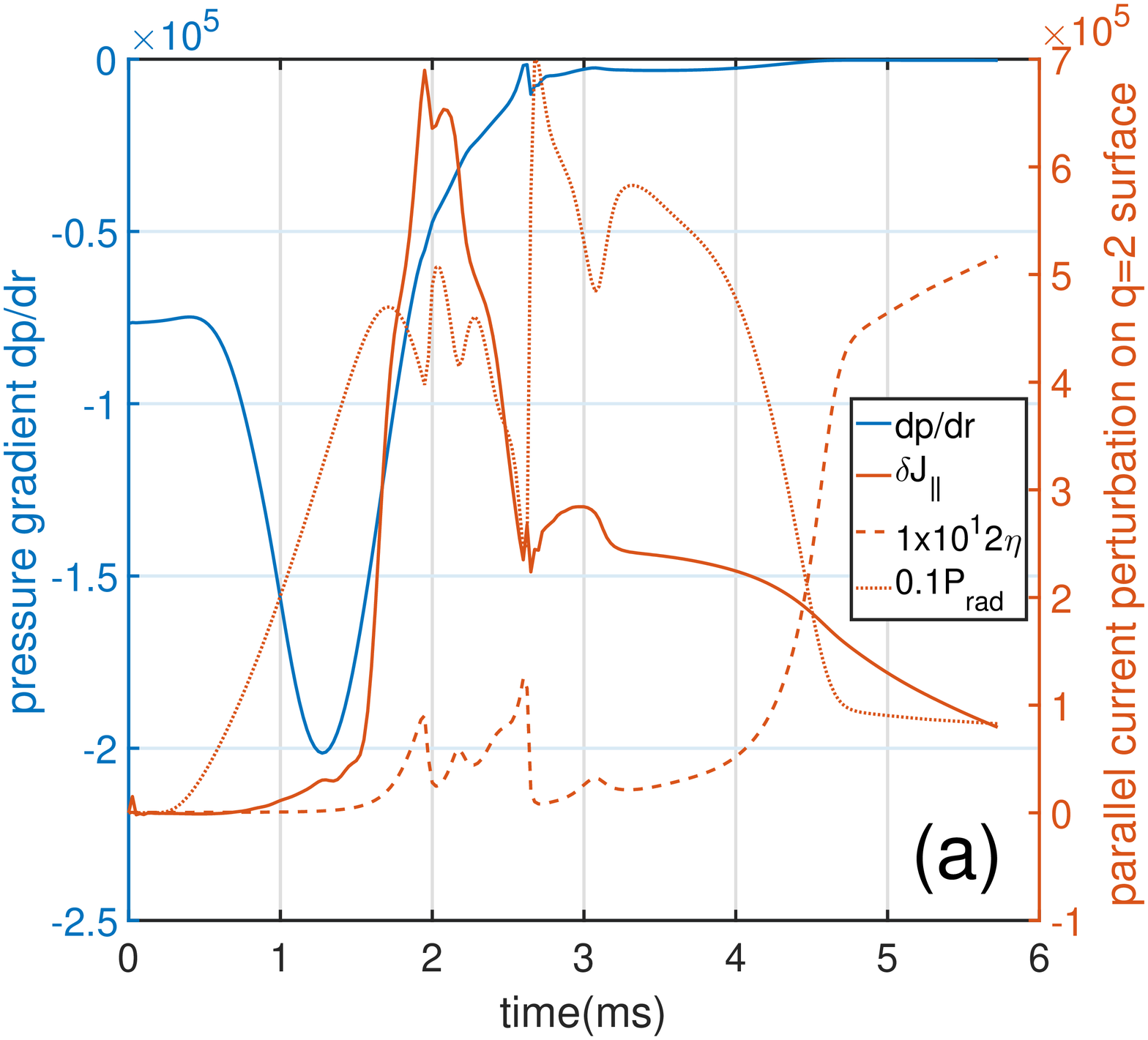}
			\includegraphics[width=0.7\textwidth,height=0.42\textheight]{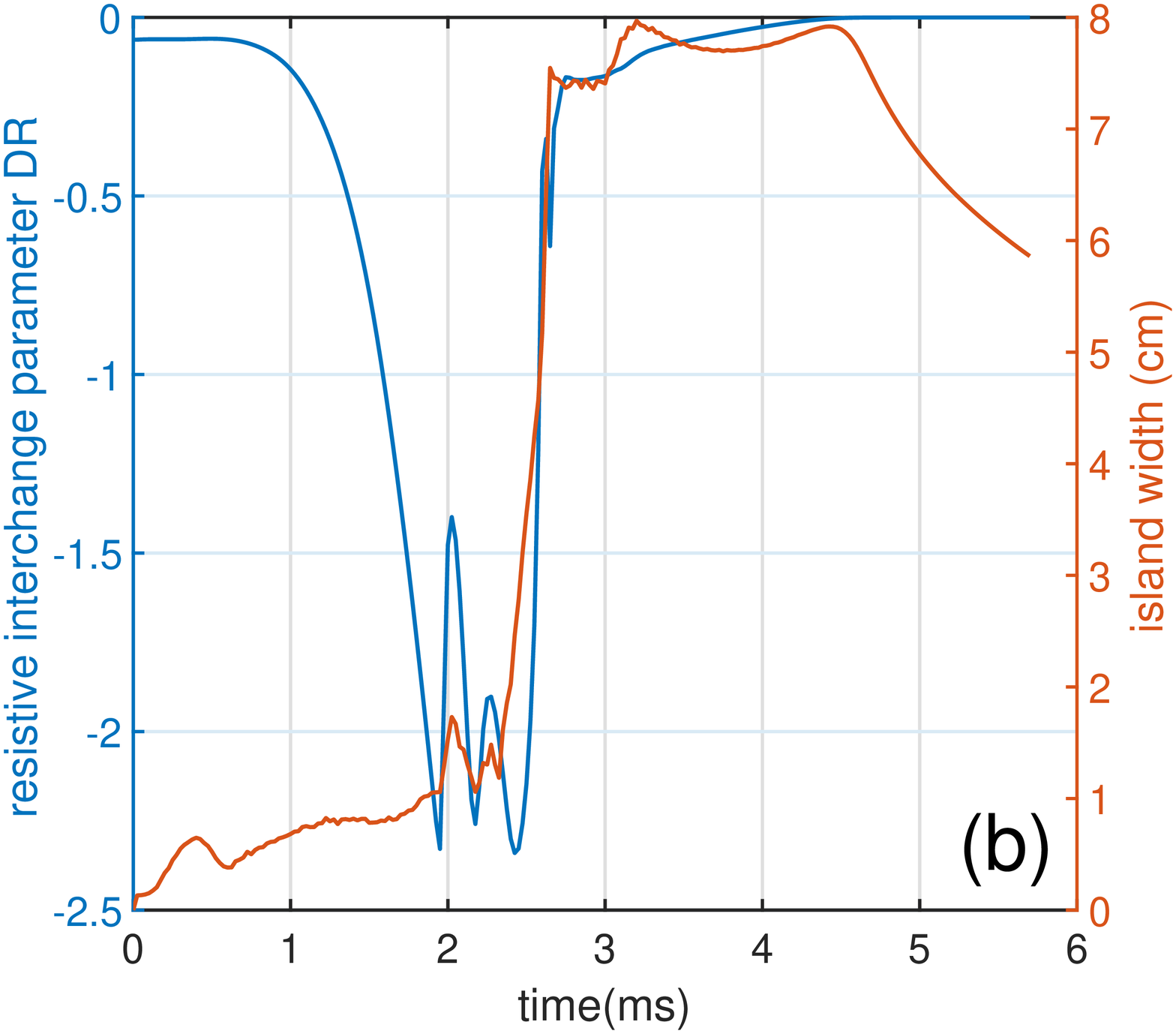}
		\end{center}
		\caption{(a) Flux-surface-averaged pressure gradient $\left\langle dp/dr\right\rangle$ (blue solid line), perturbed parallel current $\left\langle \delta J_{\parallel}\right\rangle$ (orange solid line), plasma resistivity $\eta$ (orange dashed line), and radiation power $P_{rad}$ (orange dotted line) on the $q=2$ rational surface, (b) the local resistive interchange parameter $D_R$ (blue solid line) and $2/1$ island width $w$ (orange solid line) as a function of time.}
		\label{DR-width}
	\end{figure}

	The TMs are driven by the parallel current perturbation as expected and confirmed from the correlation observed for the $2/1$ as well as the $3/1$ modes (Figs. \ref{deltaj-width}a-b).
    Based on previous developed analytic theories on the nonlinear resistive growth of magnetic island including the interchange effects, the following Modified Rutherford Equation (MRE) may be introduced for discussion \cite{Glasser1975,Kotschenreuther1985,Lutjens2001,arXiv_NF}
    \begin{equation}
    	\frac{1}{\eta} \frac{\partial w}{\partial t} \sim \Delta'  + \alpha_1 \delta J_{m,n}^{ps} + \alpha_2 E_{\phi_0}\delta \eta_{m,n}
    \end{equation}
    where $\Delta'$ is the tearing instability index, $\delta J_{m,n}^{ps}$ and $\delta \eta_{m,n}$ are the $(m,n)$ helical component of the perturbed PS current and resistivity respectively, and $E_{\phi_0}$ is the toroidal electric field due to loop voltage. Here the coefficients $\alpha_1$, and $\alpha_2$ may be determined from the more quantitative theory or fit from the corresponding simulation or experimental results. For the $2/1$ mode, the parallel current perturbation around the $q=2$ rational surface $\left\langle \delta J_{\parallel} \right\rangle_{q=2}$ increases rapidly at $t=1.5ms$, and the $2/1$ island width grows up thereafter at $t=2.5ms$ (Fig. \ref{deltaj-width}a).
	Particularly, the parallel current perturbation is primarily contributed from the enhanced impurity radiation, as indicated by the correlation between the local radiation power and the current perturbation $\delta J_{\parallel}$ in Fig. \ref{DR-width}(a), which is the main drive for the tearing growth. After $t=4.5ms$, the current perturbation becomes dominated by resistive diffusion due to the enhanced plasma resistivity.
	The $1.0 ms$ delay ($t=1.5-2.5ms$) in the mode growth may be due to the well-known stabilization effect from flux-averaged pressure gradient and magnetic curvature \cite{Glasser1975,Glasser1976}, which reduces the cylindrical tearing instability parameter $\Delta'$ to an effective value \cite{Lutjens2001}, i.e. $\Delta_{eff}' = \Delta' + \sqrt{2}\pi^{3/2} D_R/w_d$. Here $w_d$ is the finite thermal diffusion length scale \cite{RichardFitzpatrick1995}, and $D_R \approx \frac{\epsilon^2_s\beta_p}{s} \frac{L_q}{L_p} \left( 1-\frac{1}{q^2} \right)$ is the resistive interchange parameter with $\epsilon_s=r_s/R_0, \beta_p=2\mu_0p/B_p^2, L_q=q/q'=r_s/s$, and $L_p=p/p'$, which is usually negative in a tokamak with monotonic safety factor $q$ profile \cite{Glasser1975}.
	This derives from the fact that perturbed pressure leads to a parallel current perturbation outside the resistive layer through magnetic curvature, which contributes to the jump in the logarithmic derivative, i.e. the $\Delta'$.
	From Fig. \ref{DR-width}(a), it is clear that the local pressure gradient is enhanced around $1.5ms$ and decreases towards zero right before the mode growth ($t=2.5ms$).	
	More importantly, the absolute value $|D_R|$ indeed increases to a larger value right before the mode growth, which represents the stabilization effect, and drops rapidly once the mode begins to grow (Fig. \ref{DR-width}b).
	With the impurity inward penetration, the current perturbations move along with the cold front and cross different rational surfaces to trigger TMs with multiple helicities.

	\section{The interaction between the induced TM and impurity}
	\label{Section:4}
	\subsection{The rotation of magnetic islands}
\begin{figure}[ht]
		\begin{center}
			\includegraphics[width=1.0\textwidth,height=0.3\textheight]{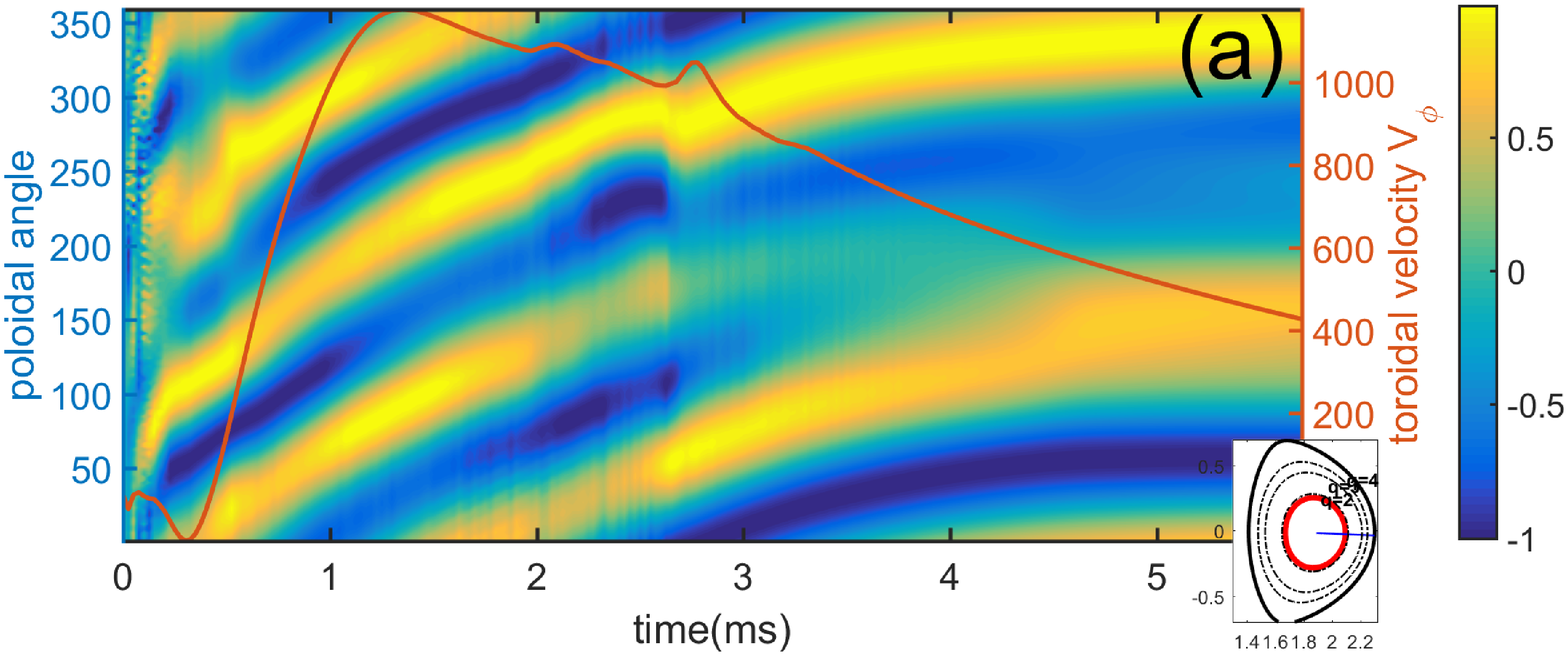}
			\includegraphics[width=1.0\textwidth,height=0.3\textheight]{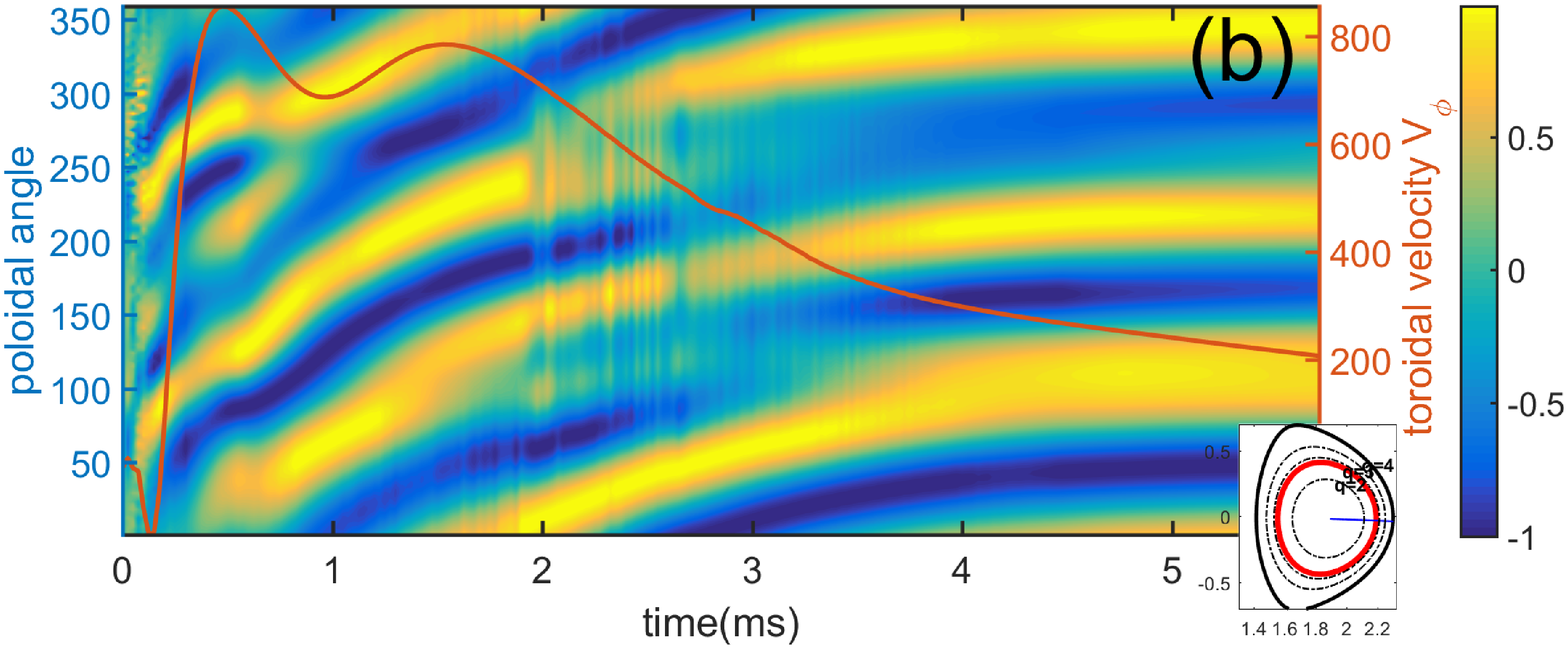}
		\end{center}
		\caption{Flux-surface-averaged toroidal velocity $V_{\phi}$ (orange solid line) and normalized $n=1$ normal component of perturbed magnetic field $B_{r,n=1}$ (flushed color), which are measured on the $q=m/n$ rational surface (denoted as the red circle in the sketch) for (a) the $2/1$ mode and (b) the $3/1$ mode as a function of time.}
		\label{vphi-br}
	\end{figure}

	A stationary equilibrium plasma framework is adopted in the simulations, i.e. $V_0 = 0 m/s$, to study the impurity effect on the rotation of magnetic islands (Fig. \ref{vphi-br}). The small perturbed velocity at the beginning of simulation is caused by the impurity injection, then it increases rapidly to its peak and decays slowly thereafter. The mode frequency of the $n=1$ normal component of the perturbed magnetic field $B_r$ is slowly damped from beginning, and the magnetic islands become almost stationary after $t \gtrsim 4ms$. The toroidal rotation frequency on the $q=2$ surface is approximate twice that on the $q=3$ surface, whereas the mode frequencies on these two surfaces are almost same during $t=2.0-4.0ms$ before locking. Such a mode excitation with finite frequency and the subsequent gradual damping and eventual locking is also observed in experiments,
	however, such a mode locking is previously attributed to the electromagnetic torque braking due to the error field from the tungsten protector limiter \cite{XuMing2022}.
	The local magnetic perturbation on the $q=2$ surface shows the initial dominant poloidal mode number $m=3$ before $t=2.5ms$, which is later replaced by the $m=2$ poloidal component only after the excitation of the $2/1$ mode by the arrival of the impurity radiation peak on the $q=2$ surface (Fig. \ref{vphi-br}a).

	The island rotation amplitude can be measured by the integral of the perpendicular vortex associated with the mode in the poloidal plane, which decreases towards zero gradually (Fig. \ref{vortex}a). Meanwhile, the impurity penetration front can be indicated by the local enhanced electron density peak, which sweeps inward in a step-wise manner across rational surfaces. The mode rotation amplitude rapidly shoots to its peak value in the beginning when the impurity is localized in the bottom region with a strong up-down asymmetric distribution, then gradually slows down as the impurity penetrates inward along with more uniform toroidal and poloidal spreading. This simulation results agree the experimental observation that the modes can be locked following the redistribution of the low-Z impurity concentration \cite{XuMing2022}, despite the fact that there is no error or external magnetic field at the perfectly conducting wall boundary in simulations. Similarly, J-TEXT experiments demonstrate strong correlation between the $2/1$ tearing mode rotation and impurity distribution as well \cite{Li_2020,Tong_2019}.

	\subsection{The effect from higher-$n$ modes}
\begin{figure}[ht]
	\begin{center}
		\includegraphics[width=0.45\textwidth,height=0.3\textheight]{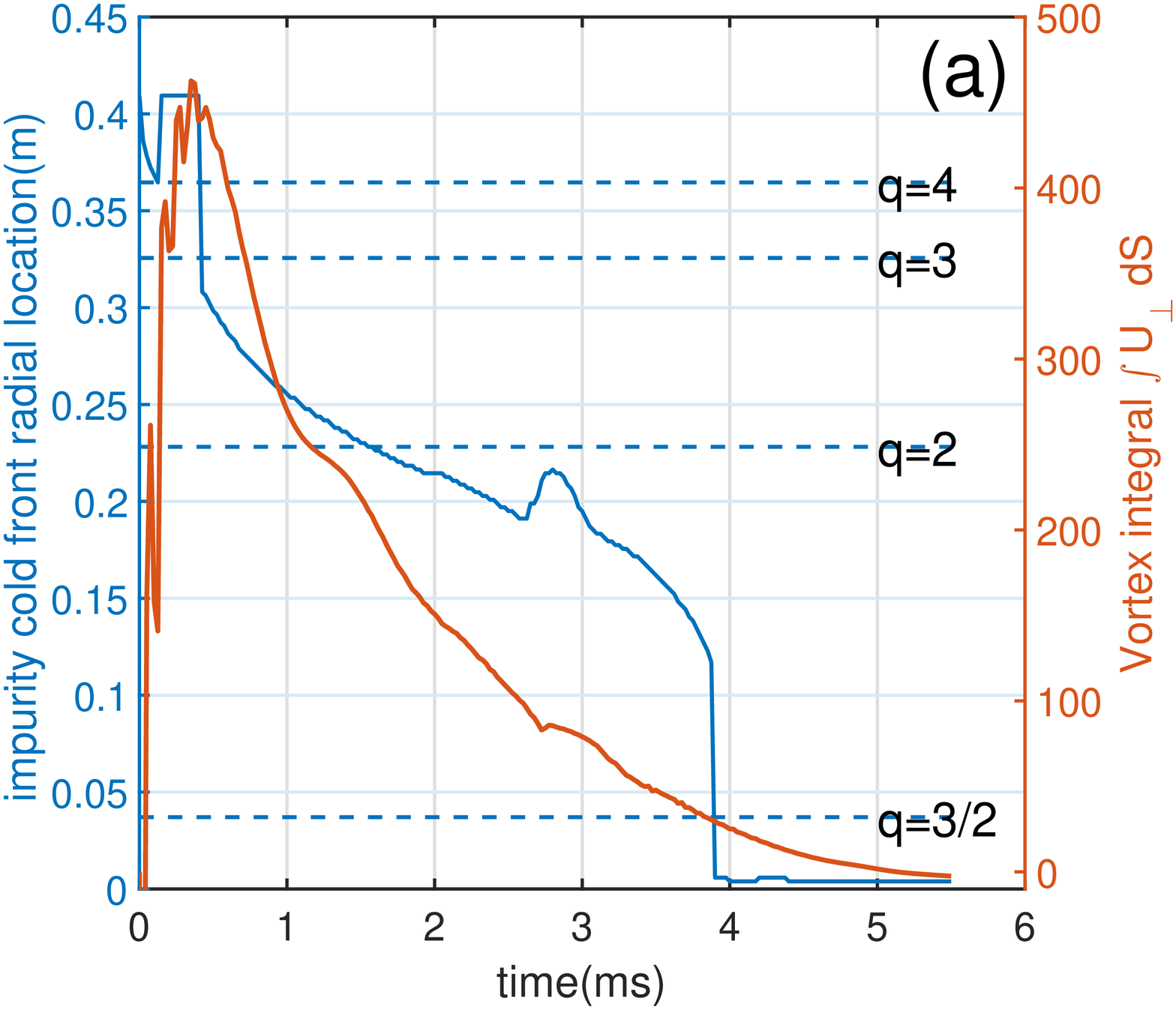}
		\includegraphics[width=0.45\textwidth,height=0.3\textheight]{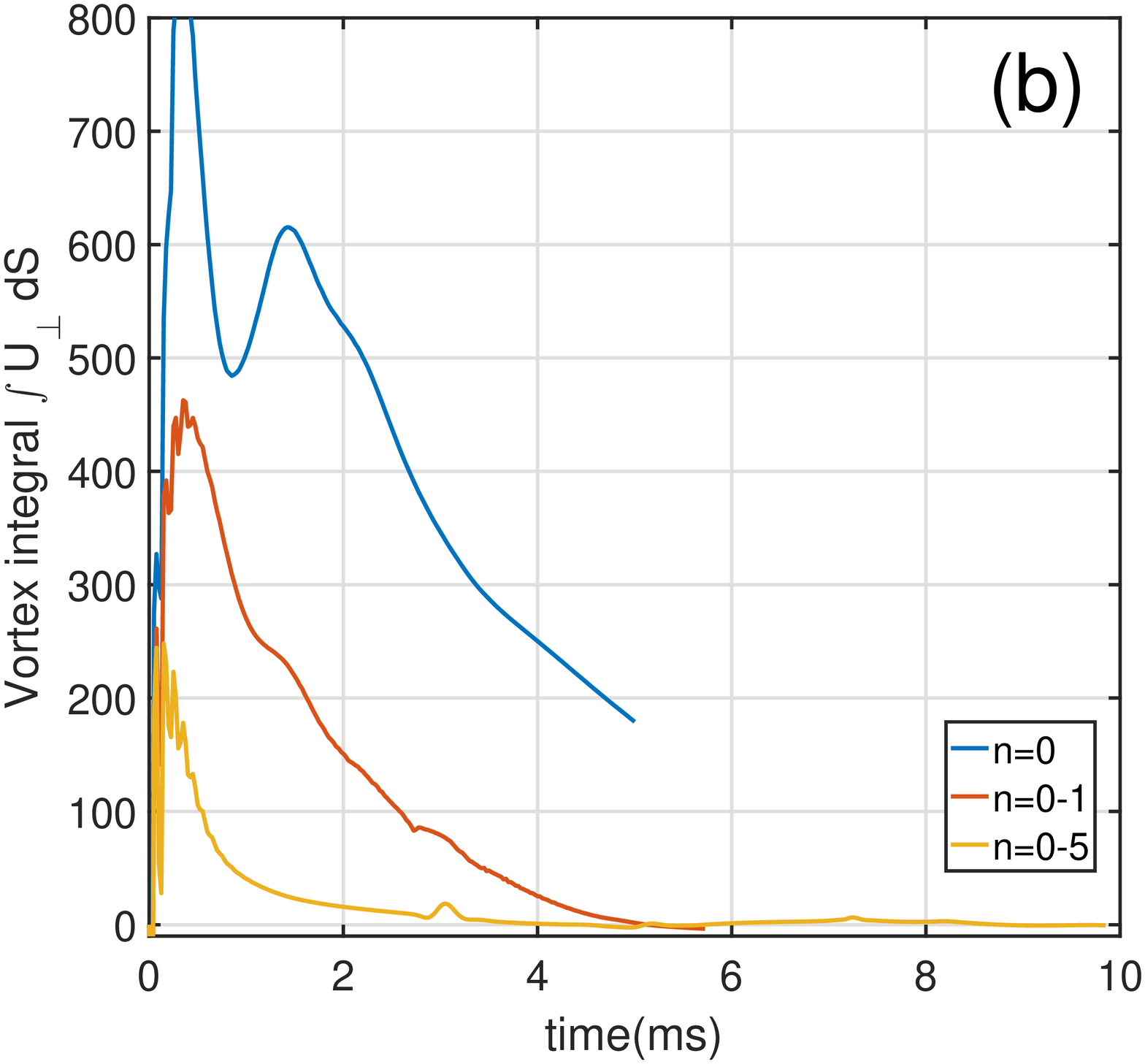}
		\includegraphics[width=0.45\textwidth,height=0.3\textheight]{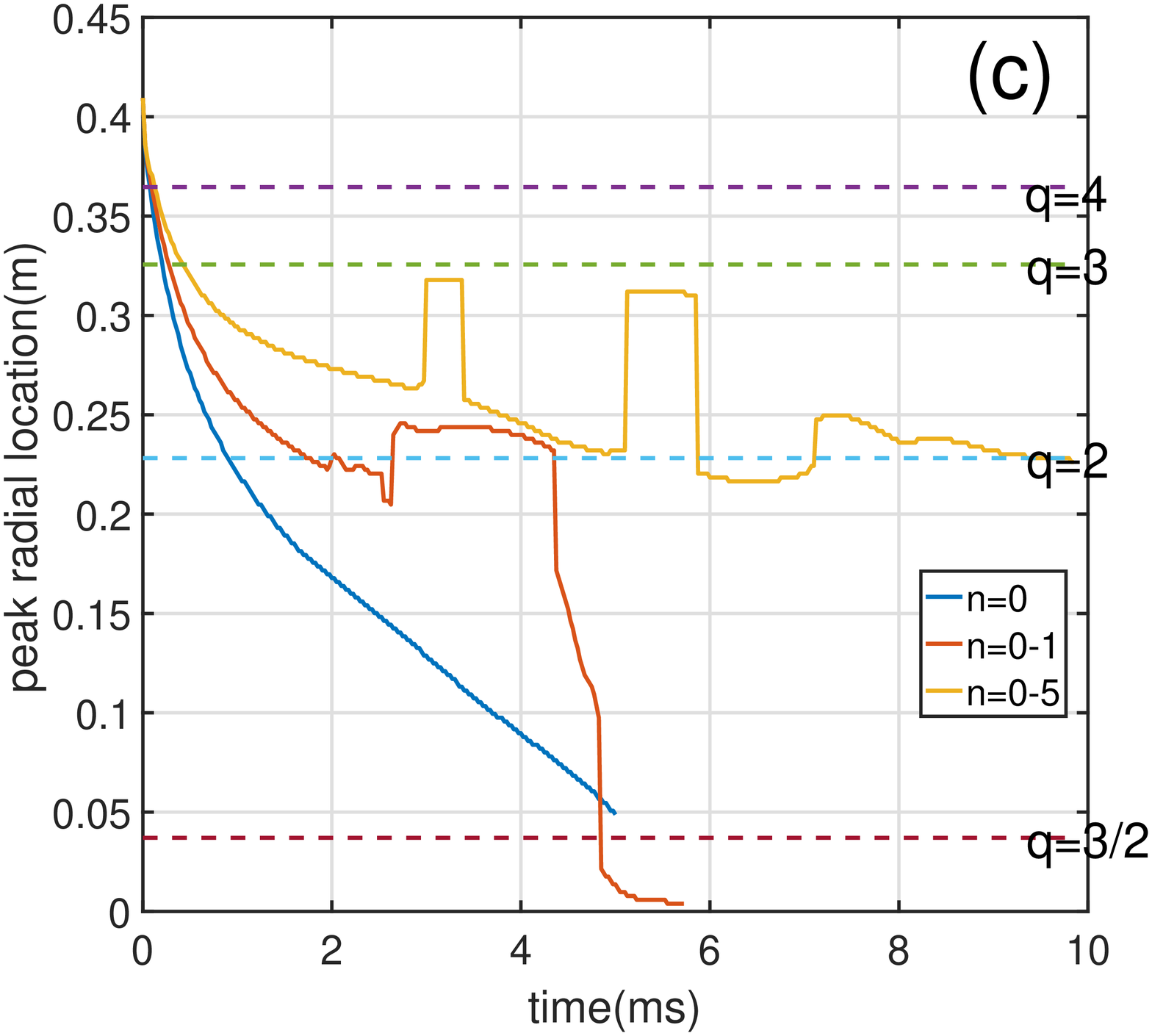}
	\end{center}
	\caption{(a) The integral of perpendicular vortex $\int U_{\perp} dS = \int \nabla_{\perp} V dS$ in the poloidal plane (orange solid line) and radial location of the perturbed electron density peak $max(\delta n_e)$ (blue solid line) as a function of time. (b) The integral of perpendicular vortex $\int U_{\perp} dS = \int \nabla_{\perp} V dS$ in the poloidal plane, and (c) the radial location of impurity radiation peak as a function of time for cases with different inclusions of toroidal mode numbers.}
	\label{vortex}
\end{figure}

	The inclusion of higher-$n$ modes in simulation accelerates the mode rotation drop towards zero and the subsequent stationary state in the simulation (Fig. \ref{vortex}b). This suggests that the high-$n$ helical structures may be able to introduce additional braking effects, likely through the electromagnetic torques from the magnetic island chains, as well as the coupling and overlapping of magnetic islands on the neighbouring rational surfaces.

	Higher-$n$ modes also significantly impede the impurity inward penetration and as a consequence the radiation peak stays longer upon the $q=2$ surface (Fig. \ref{vortex}c), which could be due to a combined effects from the stochastic field \cite{Hu_2021} and the $2/1$ magnetic island itself \cite{Izzo2017}. By contrast, in the simulation case with the $n=0$ component only, the impurity front almost directly penetrates into the central region in absence of magnetic islands.
	This agrees with the observations that during an MGI experiment, the impurity penetration usually stops along the $q=2$ surface \cite{Reux_2010,Hollmann2007}, and such an agreement highlights the critical roles of the higher-$n$ modes in the impurity penetration process.

	\section{Summary and discussion}
	\label{Section:5}
	The successive formation of tearing modes observed during an impurity penetration process on EAST has been well reproduced in our 3D resistive MHD simulations using the NIMROD code with good agreement on several main features.
	The $4/1,3/1$ and $2/1$ TMs grow in sequence after the arrival of impurity radiation peak on the corresponding rational surfaces, and the island rotation slows down gradually with the impurity accumulation.
	The current perturbations induced by the impurity penetration is found to be responsible for the island growth. The perturbed Pfirsch-Schl\"{u}ter current due to enhanced pressure gradient perturbation shows its stabilization effect whereas the radiation enhanced current perturbation predominately contributes to driving of the tearing instability. After the island saturation, the current perturbation is mainly affected by the plasma resistivity due to the radiative cooling.
	Higher-$n$ modes are found to introduce braking effects on the island rotation, and more importantly the impedance to the impurity inward penetration.			
	Whereas this work demonstrates the causal relation between the current perturbation induced by impurity radiation and the magnetic island growth, and in particular the roles of the higher-$n$ modes, more quantitative model and analyzes are to be developed in future work.

	\section{Acknowledgments}
	We are grateful for the supports from the NIMROD team. This work was supported by the National Magnetic Confinement Fusion Program of China (Grant No. 2019YFE03050004), the National Natural Science Foundation of China (Grant Nos. 11775221 and 51821005), the Fundamental Research Funds for the Central Universities at Huazhong University of Science and Technology (Grant No. 2019kfyXJJS193), and U.S. Department of Energy (Grant Nos. DE-FG02-86ER53218 and DE-SC0018001). This research used the computing resources from the Supercomputing Center of University of Science and Technology of China. 

	\section{Appendix}
	\label{appendix}

	The KPRAD module adopted in the NIMROD code is used to update the impurity charge state populations and calculate the power of impurity radiations \cite{KPRAD}, which include the background impurity radiation $P_{bg}$, the line radiation $P_{line}$, the bremsstrahlung $P_{brem}$, the 3-body recombination $P_{3-body}$, the ionization $P_{ion}$ and recombination $P_{rec}$, and the atomic physics data originates from ADAS database (URL https://www.adas.ac.uk/) \cite{ADAS}.
	
	The background impurity radiation power results from the material sputtering from the divertor or the first wall, which can be set to be beryllium $(Be)$, boron $(B)$, or carbon $(C)$
	\begin{eqnarray}
		P_{bg} = f_{z,bg} \times 10^{-13} \times n_e[m^{-3}] \times 10^{p_{bg}}  \\
		p_{bg} = \sum_{i} f_{bg(i)} \times \left(\log_{10}{T_e[keV]} \right)^{i-1}
	\end{eqnarray}
	where $f_{z,bg} = n_{Z,bg}/n_i$ is the fraction of background impurity density, $n_i$ the plasma ion density, $n_e$ the electron density, $n_{Z,bg}$ the background impurity density, $p_{bg}$ is the polynomial as a function of the electron temperature $T_e$ based on the coronal equilibrium and $f_{bg(i)}$ are the fitted coefficients of the radiation curve, in particular, the range of temperature $T_e$ only includes $2\sim20\ keV$ in the background impurity radiation \cite{KPRAD,ADAS}.
	
	The impurity line radiation power can be set to be helium $(He)$, beryllium $(Be)$, carbon $(C)$, neon $(Ne)$, or argon $(Ar)$
	\begin{eqnarray}
		P_{line,Z(c)} = 10^{p_{line}} \times 10^{-13} \times n_e[cm^{-3}] \times n_{Z,(c-1)}[cm^{-3}] \\ 
		p_{line} = \sum_{i} f_{line(i)} \times \left(\log_{10}{T_e[keV]} \right)^{i-1}
	\end{eqnarray}
	where $Z$ is the atomic number and $c=0-Z$ is the impurity charge state, the line radiation of impurity charge state $c$ correlates to the density of its former charge state $n_{Z,(c-1)}$, $p_{line}$ is the polynomial as a function of $T_e$ and $f_{line(i)}$ are the fitted coefficients of the line radiation curve \cite{KPRAD,ADAS}.
	
	The bremsstrahlung radiation power
	\begin{eqnarray}
		P_{brem} = 1.69 \times 16^{-32} \times n_e^2[cm^{-3}] \times \sqrt{T_e[eV]} \times Z_{eff}  \\
		Z_{eff} = 1 + \sum_{c}^{Z} \frac{(c^2-c) \times n_c[cm^{-3}]}{n_e [cm^{-3}]}
	\end{eqnarray}
	where $Z_{eff}$ is the effective charge state number and $n_c$ is the impurity density of different charge state.
	
	The 3-body recombination radiation power
	\begin{equation}
		P_{3-body,Z(c)} = 8.75 \times 10^{-39} \times n_e^2[cm^{-3}] \times c^3 \times T_e^{-4.5} [eV]
	\end{equation}
	which is proportional to $T_e^{-4.5}$ and becomes important only at $T_e \simeq 1 eV$.
	
	The ionization radiation power
	\begin{eqnarray}
		P_{ion,Z(c)} = 1.6\times 10^{-19} \times R_{ion,Z(c)} \times n_{Z,(c-1)}[cm^{-3}] \times E_{ion,Z(c)}[eV] \\
		R_{ion,Z(c)} = n_e[cm^{-3}] \times 10^{\sum_{i} f_{ion(i)} \times\left( \log_{10}{T_e[eV]} \right)^{i-1}}
	\end{eqnarray}
	and the recombination radiation power
	\begin{eqnarray}
		P_{rec,Z(c)} = 1.6 \times 10^{-19} \times R_{rec,Z(c)} \times n_{Z,(c)}[cm^{-3}] \times\left( E_{ion,Z(c)}[eV] + T_e \right) \\
		R_{rec,Z(c)} = 5.2 \times 10^{-14} \times n_e[cm^{-3}] \times \left( c+1\right) \times \sqrt{\frac{E_{ion,Z(c)}[eV]}{T_e[eV]}} \times f_{rec} \\
		f_{rec} = 0.43 + 0.5 \times \log_{10}{\frac{E_{ion,Z(c)}[eV]}{T_e[eV]}} + 0.469\times\left(\frac{E_{ion,Z(c)}[eV]}{T_e[eV]} \right)^{-1/3}
	\end{eqnarray}
	where $E_{ion,Z(c)}$ is the ionization energy of impurity charge state $c$, $R_{ion,Z(c)}$ is the ionization rate and $f_{ion(i)}$ are the fitted coefficients of the polynomial for the ionization radiation curve, $R_{rec,Z(c)}$ is the recombination rate as a function of electron density $n_e$ and temperature $T_e$ \cite{KPRAD,ADAS}. The ionization is closely associated with the recombination and note that the ionization radiation of impurity charge state $c$ correlates to the density of its former charge state $n_{Z,(c-1)}$. Besides, the ionization and the recombination rates $R_{ion,Z(c)}$ and $R_{rec,Z(c)}$ are used to update each impurity charge state density respectively in the source terms $S_{ion}$ and $S_{rec}$ of the continuity equation at every time step.

	\newpage
	\section{Reference}
	
	\bibliographystyle{iopart-num}
	\bibliography{eastexptm}

	\newpage

	\newpage

	\newpage

	\newpage

	\newpage

	\newpage

	\newpage

	\newpage

\newpage

\end{document}